\documentclass[twocolumn,superscriptaddress,reprint,nofootinbib,prd]{revtex4-1}%
\usepackage{amssymb}
\usepackage{color}
\usepackage{graphicx}
\usepackage{multirow}
\usepackage{dcolumn}
\usepackage{bm}
\usepackage[header,title,page,titletoc]{appendix}
\usepackage[utf8]{inputenc}
\usepackage{amsmath}
\usepackage{bbm}
\usepackage{amsfonts}%
\usepackage[bookmarks,colorlinks=false]{hyperref}
\newlength{\wordlength}

\newlength{\onewordlength}

    \newcommand{\ba}{\begin{eqnarray}}
    \newcommand{\ea}{\end{eqnarray}}
    \newcommand{\be}{\begin{equation}}
    \newcommand{\ee}{\end{equation}}

\newcommand{\bn}{{\bf n}}

\newcommand {\bk} {{\mathbf k}}

\newcommand {\bp} {{\mathbf p}}

\newcommand{\bx}{{\bf x}}

\newcommand{\calC}{{\mathcal C}}

\newcommand{\calP}{{\mathcal P}}

\newcommand{\calO}{{\mathcal O}}

\newcommand{\calM}{{\mathcal M}}
\newcommand{\calZ}{{\mathcal Z}}

\newcommand{\calV}{{\mathcal V}}

\begin{document}

\title{A coupled-channel lattice study on the resonance-like structure $Z_c(3900)$}

\author{Ting Chen}
\affiliation{%
School of Physics, Peking University, Beijing 100871, China
}%

\author{Ying Chen}
\affiliation{%
Institute of High Energy Physics, Chinese Academy of Sciences, Beijing 100049, China\\
School of Physics, University of Chinese Academy of Sciences, Beijing 100049, China
}%

\author{Ming Gong}
\affiliation{%
Institute of High Energy Physics, Chinese Academy of Sciences, Beijing 100049, China\\
School of Physics, University of Chinese Academy of Sciences, Beijing 100049, China
}%

\author{Chuan Liu}%
\email[Corresponding author. Email: ]{liuchuan@pku.edu.cn}
\affiliation{%
School of Physics and Center for High Energy Physics, Peking
University, Beijing 100871, China\\
Collaborative Innovation Center of Quantum Matter, Beijing 100871, China
}%

\author{Liuming Liu}
\affiliation{Institute of Modern Physics,Chinese Academy of Sciences, Lanzhou 730000, China\\
University of Chinese Academy of Sciences, Beijing 100049, China}

\author{Yu-Bin Liu}
\affiliation{%
School of Physics, Nankai University, Tianjin 300071, China
}

\author{Zhaofeng Liu}
\affiliation{%
Institute of High Energy Physics, Chinese Academy of Sciences, Beijing 100049, China\\
School of Physics, University of Chinese Academy of Sciences, Beijing 100049, China
}

\author{Jian-Ping Ma}
\affiliation{%
Institute of Theoretical Physics, Chinese Academy of Sciences, Beijing 100190, China
}

\author{Markus Werner}
\affiliation{%
 Helmholtz--Institut f\"ur Strahlen-- und Kernphysik and Bethe Center for Theoretical Physics,\\
 Universit\"at Bonn, D--53115 Bonn, Germany
}

\author{Jian-Bo~Zhang}
\affiliation{%
Department of Physics, Zhejiang University, Hangzhou 311027, China
}

 \collaboration{CLQCD Collaboration}

 \begin{abstract}
 In this exploratory study, near-threshold scattering of $D$ and $\bar{D}^*$ meson is investigated using
 lattice QCD with $N_f=2+1+1$ twisted mass fermion configurations.
 The calculation is performed within the coupled-channel L\"uscher's finite-size
 formalism. The study focuses on the channel
 with $I^G(J^{PC})=1^+(1^{+-})$ where the resonance-like structure $Z_c(3900)$
 was discovered. We first identify the most relevant two channels of the problem
 and the lattice study is performed within the two-channel scattering model.
 Combined with a two-channel Ross-Shaw theory, scattering parameters are extracted from the
 energy levels by solving the generalized eigenvalue problem.
 Our results on the scattering length parameters suggest that,
 at the particular lattice parameters that we studied,
 the best fitted parameters do not correspond to a peak behavior in
 the elastic scattering cross section near the threshold. Furthermore, within the zero-range
 Ross-Shaw theory, the scenario of a narrow resonance close to the threshold
 is disfavored beyond $3\sigma$ level.
 \end{abstract}

 \maketitle



 \section{Introduction}

 In the past decade or so, various exotic hadronic resonance-like structures
 have been witnessed by numerous experimental groups.
 These structures, due to their unknown nature, are generally called $XYZ$ particles.
 More interesting ones are the charged structures that have been discovered
 in both charm and bottom sectors. These structures necessarily bear a
 four valence quark structure $\bar{Q}q\bar{q}'Q$ with $Q$ being a heavy-flavor
 quark while $q$ and $q'$ being the light-flavored quark. For different light flavors,
 these are charged objects. Another interesting feature is that they tend to appear close
 to the threshold of two heavy mesons with valence structure $\bar{Q}q$ and $\bar{q}'Q$.
 The physical nature of these structures have
 been contemplated and discussed in many phenomenological studies.
 For example, they could be shallow bound states of the two mesons due to residual color interactions
 or some genuine tetraquark objects.
 However, even after so many phenomenological  studies, the nature of many of these states remains obscure.
 A typical example is the structure $Z_c(3900)$, which will be the main topic of this paper.
 It was first discovered by BESIII~\cite{Ablikim:2013mio} and
 Belle~\cite{Liu:2013dau} and soon also verified by CLEO collaborations~\cite{Xiao:2013iha}.
 The nature of $Z_c(3900)$ remains in debate.
 For a recent review on these matters, see e.g. Ref.~\cite{Liu:2016ahj,Guo:2017jvc}.
 It is therefore highly desirable that non-perturbative methods like lattice QCD
 could provide some information on the nature of these states.

 Quite contrary to many phenomenological studies, lattice studies on these states
 are still relatively scarce. For the state $Z_c(3900)$,
 it is readily observed that the invariant mass of the structure is close to the $D\bar{D}^*$ threshold,
 naturally suggesting a shallow molecular bound state formed by the two corresponding charmed mesons.
 To further investigate this possibility, the interaction between
 $D^*$ and $D$ mesons near the threshold becomes crucial.

 A lattice study was performed by S.~Prelovsek et al. who investigated the energy levels
 of the two charmed meson system in the channel where $Z_c$ appearing
 in a finite volume~\cite{Prelovsek:2014swa}.
 They used quite a number of operators, including two-meson operators
 in the channel of $J/\psi\pi$, $D\bar{D}^*$ etc. and even tetraquark operators.
 However, they discovered no indication of extra new energy levels apart from the
 almost free scattering states of the two mesons. Taking $D\bar{D}^*$ as
 the main relevant channel, CLQCD utilized single-channel L\"uscher scattering formalism~\cite{luscher86:finitea,luscher86:finiteb,luscher90:finite,luscher91:finitea,luscher91:finiteb} to
 tackle the problem and also found slightly repulsive interaction between the two charmed mesons~\cite{Chen:2014afa,Chen:2015jwa}.
 Therefore, it is also unlikely for them to form bound states.
 A similar study using staggered quarks also finds no clue for the existence of the state~\cite{Lee:2014uta}.
 Thus, the above mentioned lattice studies, whether it is inspecting the energy levels alone or
 utilizing single-channel L\"uscher approach,  have found no clue for the existence of
 a $D\bar{D}^*$ bound state.

 On the other hand, starting around 2015, HALQCD studied the problem using
 the so-called HALQCD approach~\cite{Ishii:2006ec} which is rather different from L\"uscher's adopted
 by other groups. They claimed that this structure can be reproduced and it is not a usual bound state or resonance,
 but rather, a structure formed due to strong cross-channel interactions, see Ref.~\cite{Ikeda:2016zwx,Ikeda:2017mee} and references therein.
 So far, this scenario is only seen in the HALQCD approach but not within the L\"uscher-type approach.
 In fact, the multi-channel L\"uscher formula is already known~\cite{chuan05:2channel,Liu:2005kr,Lage:2009zv,Doring:2011vk,Doring:2012eu}.
 Using this multi-channel L\"uscher approach,
 Hadron Spectrum Collaboration has successfully studied various coupled-channel scattering
 processes involving light mesons~\cite{Dudek:2014qha,Wilson:2014cna,Wilson:2015dqa,Briceno:2016mjc}.
 It is therefore tempting for us to verify/falsify the cross-channel interaction scenario
 suggested by HALQCD in a coupled-channel L\"uscher approach.
 More importantly, the state $Z_c$ does have many coupled decay channels
 and if the coupled channel effects are important, then inspecting the energy levels alone
 or performing only a single-channel scattering study will not be adequate to understand
 the nature of these structures. Therefore, in this exploratory study, we aim to
 make a step towards the multi-channel lattice computation using the
 coupled-channel L\"uscher approach.

 It is well-known that, within single-channel L\"uscher's approach, energy levels are in one-to-one correspondence
 with the scattering phase. However, in a two-channel situation,
 the $S$-matrix is characterized by 3 parameters, all of which are functions
 of the scattering energy. Therefore, one needs to re-parameterize the
 $S$-matrix elements in terms of a number of constant parameters so as to pass from the energy levels
 to the scattering phases. One possible choice is to utilize the so-called $K$-matrix parameterization
 adopted by the Hadron Spectrum Collaboration in their studies of light meson coupled-channel scattering.
 In this work, however, since we are only interested in the energy
 region very close to the threshold,
 we will be using the so-called multi-channel effective range expansion developed
 long time ago by Ross and Shaw~\cite{Ross:1960len,Ross:1961ere}.
 We will call this the Ross-Shaw theory.
 The difficulty with the multi-channel approach lies in the fact that,
 the number of parameters needed to parameterize the $S$-matrix grows quadratically fast
 when the number of channels is increased.
 Therefore, based on the experimental facts
 and also hints from the HALQCD study, we attempt to study this problem
 within the two-channel L\"uscher approach. This could be viewed as the
 first step beyond the single-channel approximation using L\"uscher's formalism.
 To be more specific, we will single out two most relevant channels for $Z_c(3900)$:
 $J/\psi\pi$ and $D\bar{D}^*$, the first being
 the discovery channel for $Z_c(3900)$ and the second has shown to be
 the dominant channel by BESIII experimental data~\cite{Ablikim:2015gda}.
 Quite similar to the single channel effective range expansion which
 is characterized by two real parameters, namely the scattering length $a_0$
 and the effective range $r_0$, in a two-channel situation, one needs 5 real parameters to describe
 the so-called Ross-Shaw matrix $M$: 3 for the scattering length
 matrix and 2 for the effective range parameters.
 If one would like to go beyond two channels, these numbers go up
 rather quickly. For example, for the case of three channels, the scattering length matrix has 6 real parameters
 while the effective range will add another 3, making the total number
 of parameters 9 which we think is already too many to handle.
 Thus, in this paper, using lattice QCD within
 the two-channel L\"uscher's formalism combined with
 the Ross-Shaw effective range expansion, our aim is to check whether
 the cross-channel interaction scenario as suggested by the HALQCD can be realized or not.

 This paper is organized as follows. In Section~\ref{sec:method}, we briefly
 outline the computational strategies in L\"uscher's formalism and
 review the Ross-Shaw effective range expansion that we used to parameterize
 out $S$-matrix elements. In particular, we discuss the conditions that
 should be satisfied in order to have a narrow resonance close to the threshold.
 In section~\ref{sec:operators}, interpolating operators are introduced from which
 correlation matrices can be computed. We also outline how the most relevant
 two channels are determined from these correlation matrices.
 In section~\ref{sec:simulation_details}, simulation
 details are given and the results for the single- and two-meson systems
 are analyzed. By applying the two-channel L\"uscher's
 formula, parameters that determines the Ross-Shaw $M$-matrix are extracted
 and the physics implied by them are also discussed.
 In Section~\ref{sec:conclude}, we will conclude with some general remarks.

 \section{Strategies for the computation}
 \label{sec:method}

 In this section, we will outline the computational strategies for
 the computation described in this paper.
 We start by reviewing the ingredients in L\"uscher's formalism
 with the focus on the multi-channel version. Then we will describe
 the multi-channel effective range expansion developed
 by Ross and Shaw. Relation of the scattering cross section
 with respect to the parameters in Ross and Shaw theory
 are also outlined which will help us understand the
 meanings of these parameters.

 \subsection{Multi-channel L\"uscher formula}

 Within the original single-channel L\"uscher formalism,
 the exact energy eigenvalue of a two-particle system
 in a finite box of size $L$ is
 related to the elastic scattering phase of the two particles
 in the infinite volume. For simplicity, we will only consider
 the center of mass frame with periodic boundary conditions applied
 in all three spatial directions. Consider two interacting bosonic particles,
 which will be called mesons in the following, with mass $m_1$ and $m_2$ enclosed in a cubic box of size $L$.
 The spatial momentum $\bk$ of any particle is quantized according to:
 \be
 \label{eq:free_k}
 \bk=\left({2\pi\over L}\right)\bn\;,
 \ee
 with $\bn$ being a three-dimensional integer.
 The exact energy of the two-particle
 system in this finite volume is denoted as: $E_{1\cdot2}(\bk)$.
 This can be obtained from lattice simulations from
 appropriate correlation functions.
 Defining a variable $\bar{\bk}^2$ via:
 \be
 E_{1\cdot 2}(\bk)=\sqrt{m^2_1+\bar{\bk}^2}
 +\sqrt{m^2_2+\bar{\bk}^2}\;,
 \ee
 which would be the energy of two freely moving particles in infinite volume
 with mass $m_1$ and $m_2$ bearing three-momenta $\bk$ and $-\bk$, respectively.
 It is also convenient to further define a variable $q^2$ as:
 \be
 q^2=\bar{\bk}^2L^2/(2\pi)^2\;,
 \ee
 which differs from $\bn^2$ due to the interaction between the two particles.
 Single-channel L\"uscher's formula gives us a direct relation
 of $q^2$ and the elastic scattering phase shift $\tan\delta(q)$
 in the infinite volume:~\cite{luscher91:finitea}
 \be
 \label{eq:luscher_cube_single}
 q\cot\delta_0(q)={1\over \pi^{3/2}}\calZ_{00}(1;q^2)\;,
 \ee
 where $\calZ_{00}(1;q^2)$ is the zeta-function which
 can be evaluated numerically once its argument $q^2$ is given.
 This relation can also address the issue of bound states
 which is related to the phase shift $\delta(k)$ analytically
 continued below the threshold, see e.g. Ref.~\cite{luscher91:finitea}.

 For the two-channel case, the $S$-matrix now becomes a $2\times 2$ matrix
 in channel space. For example, for strong interaction, the $S$-matrix is
 usually expressed as,
 \be
 \label{eq:S2-canonical}
 S=\left[\begin{array}{cc}
 S_{11} & S_{12}\\
 S_{12} & S_{22} \end{array}\right]=\left[\begin{array}{cc}
 \eta e^{2i\delta_1} & i\sqrt{1-\eta^2}e^{i(\delta_1+\delta_2)}\\
 i\sqrt{1-\eta^2}e^{i(\delta_1+\delta_2)} & \eta e^{2i\delta_2}
 \end{array}\right]\;,
 \ee
 where $\delta_1$ and $\delta_2$ are scattering phases in channel 1 and 2 respectively
 and $\eta\in[0,1]$ is called the inelasticity parameter. Note that, all three
 parameters, $\delta_1$, $\delta_2$ and $\eta$ are functions of the energy.
 It is known that below the threshold, $\eta=1$ so that the coupling between
 the two channels are turned off kinematically.

 The two-channel L\"uscher formula now takes the form,
 \be
  \label{eq:luscher_cube_two}
  \left|\begin{array}{cc}
  \frac{\calM(k^2_1)+i}{\calM(k^2_1)-i}-S_{11} & \sqrt{\frac{k_2m_2}{k_1m_1}}S_{12}\\
  \sqrt{\frac{k_2m_2}{k_1m_1}}S_{12} & \frac{\calM(k^2_2)+i}{\calM(k^2_2)-i}-S_{22}
  \end{array}\right|=0\;.
 \ee
 The function $\calM$ involves zeta-functions and
 the arguments $k^2_1$ and $k^2_2$ in this formula are related to the
 energy of the two-particle system via,
 \be
 \label{eq:erengy_dual_experssion}
 E=\frac{k^2_1}{2m_1}=E_T+\frac{k^2_2}{2m_2}\;,
 \ee
 with $E_T$ being the threshold energy.
 \footnote{Although this expression is written in non-relativistic form,
 it is easily modified to the relativistic form. As shown by L\"uscher, this
 is legitimate if we neglect the polarization effects (exponentially suppressed
 as $e^{-m_\pi L}$) which is what we always assume to be the case.}
 To be specific, $E_T=m_D+m_{D^*}-(m_{J/\psi}+m_\pi)$ and $m_1$ and $m_2$
 should be taken as the reduced mass of the $J/\psi\pi$ and $D\bar{D}^*$ systems
 respectively.

 For a given partial-wave $l$,
 the cross sections above the inelastic threshold consists of elastic cross section $\sigma^{(l)}_{e}$
 and the reaction cross section $\sigma^{(l)}_{r}$. They are given by,
 \be
 \label{eq:Xsections}
 \begin{aligned}
 \sigma^{(l)}_{e} &=\frac{\pi}{k^2_1}(2l+1)|1-S^{(l)}_{11}|^2
 \;,\\
 \sigma^{(l)}_{r}&=\frac{\pi}{k^2_2}(2l+1)(1-|S^{(l)}_{11}|^2)
 \;,
 \end{aligned}
 \ee
 where $S^{(l)}_{ij}$ being the $S$-matrix elements for partial-wave $l$ and we have
 utilized the unitarity condition for the $S$-matrix.

 \subsection{Ross-Shaw theory}

 As is mentioned above, since $\delta_1$, $\delta_2$ and $\eta$
 are all functions of the energy, two-channel L\"uscher formula~(\ref{eq:luscher_cube_two})
 gives a relation among three functions. It is therefore crucial to have
 a parameterization of the $S$-matrix in terms of constants instead of functions
 and the multi-channel effective range expansion developed by Ross and Shaw~\cite{Ross:1960len,Ross:1961ere}
 serves this purpose. Here we will briefly recapitulate the major points of that theory.

 In the single-channel case, this theory is just the well-know effective range expansion
 for low-energy elastic scattering,
 \be
 \label{eq:ere_single}
 k\cot\delta(k)=\frac{1}{a_0} +\frac{1}{2} r_0k^2+\cdots\;,
 \ee
 where $\cdots$ designates higher order  terms in $k^2$
 that vanish in the limit of $k^2\rightarrow 0$.
 Therefore, in low-energy elastic scattering, the scattering length $a_0$
 and the effective range $r_0$ completely characterize the scattering process.
 Ross-Shaw theory simply generalize the above theory to the case of multi-channels.
 For that purpose, they define a matrix $M$ via
 \be
 \label{eq:M_def}
 M=k^{1/2}\cdot K^{-1} \cdot k^{1/2}\;,
 \ee
 where $k$ and $K$ are both matrices in channel space.
 The matrix $k$ is the kinematic matrix which is a diagonal
 matrix given by
 \be
 \label{eq:kinematic_matrix_def}
 k=\left(\begin{array}{cc}
 k_1 & 0\\
 0 & k_2\end{array}\right)\;,
 \ee
 and $k_1$ and $k_2$ are related to the energy $E$ via Eq.~(\ref{eq:erengy_dual_experssion}).
 The matrix $K$ is called the $K$-matrix in scattering theory
 whose relation with the $S$-matrix is given by,
 \footnote{$K$-matrix is hermitian so that $S$-matrix is unitary.}
 \be
 \label{eq:SK_relation}
 S=\frac{1+iK}{1-iK}\;.
 \ee
 Another useful formal expression for the matrix $K$ is
 \be
 K=\tan\delta\;,
 \ee
 where both sides are interpreted as matrices in channel space.
 From the above expressions, it is easily seen that $K^{-1}$ that appears
 in Eq.~(\ref{eq:M_def}) is simply the matrix $\cot\delta$ and  without
 cross-channel coupling, the $M$-matrix is also diagonal
 with entries $M\sim\mbox{Diag}(k_1\cot\delta_1,k_2\cot\delta_2)$.
 Thus, it is indeed a generalization of the single channel case
 in Eq.~(\ref{eq:ere_single}).
 In their original paper, Ross and Shaw showed that the $M$-matrix
 as function of energy $E$ can be Taylor expanded around some
 reference energy $E_0$ as,
 \be
 \label{eq:ross-shaw-ere}
 M_{ij}(E)=M_{ij}(E_0)+\frac{1}{2}R_i\delta_{ij}\left[k^2_i(E)-k^2_i(E_0)\right]\;,
 \ee
 where we have explicitly written out the channel indices $i$ and $j$.
 The matrix $M_{ij}(E_0)\equiv M^{(0)}_{ij}$ is a real symmetric matrix in channel space that
 we will call the inverse scattering length matrix;
 $R\equiv\mbox{Diag}(R_1,R_2)$ is a diagonal matrix which we shall call
 the effective range matrix. $k^2_i$ are the entries for the kinematic
 matrix defined in Eq.~(\ref{eq:kinematic_matrix_def}). Therefore, for two channels, there are
 altogether 5 parameters to describe the scattering close to some
 energy $E_0$: 3 in the inverse scattering length matrix $M^{(0)}$
 and 2 in the effective range matrix $R$. As was shown in Ref.~\cite{Ross:1961ere} in many cases,
 $R_1\simeq R_2$ and we have only $4$ parameters in this description.
 One could further reduce the number of parameters to 3 by neglecting
 terms associated with effective ranges. This is called the zero-range
 approximation~\cite{Ross:1960len}.
 For convenience, we usually take $E_0$ to be the threshold of the
 second channel. In such a case, the two-channel Ross-Shaw $M$-matrix
 looks like,
 \be
 M(E)=\left(\begin{array}{cc}
 M_{11}+\frac{R_1}{2}[k^2_1-k^2_{10}] & M_{12}
 \\
 M_{12} &  M_{22}+\frac{R_2}{2}k^2_2
 \end{array}\right)\;,
 \ee
 where $k^2_{10}\equiv k^2_1(E=E_T)$.

 It is understood that Ross-Shaw parameterization in
 Eq.~(\ref{eq:ross-shaw-ere}) is equivalent to the
 so-called $K$-matrix parameterization with two poles.
 In this $K$-matrix representation, assuming there are altogether $n$ open channels,
 the $n\times n$ $K$-matrix is parameterized as,
 \be
 \label{eq:K-matrix-poles}
 K(E)=k^{1/2}\cdot\left(\sum^n_{\alpha=1} \frac{\gamma_\alpha\otimes\gamma^T_\alpha}{E-E_\alpha}\right)
 \cdot k^{1/2}
 \;,
 \ee
 where $k$ is the kinematic matrix analogue of Eq.~(\ref{eq:kinematic_matrix_def}),
 the label $\alpha=1,2,\cdots,n$ designates the channels and each
 $\gamma_\alpha$ is a $1\times n$ real constant matrix (an $n$-component vector). It is shown
 in Ref.~\cite{Ross:1961ere} that this is equivalent to the effective range expansion~(\ref{eq:ross-shaw-ere})
 but the parameters are more flexible. In particular,
 $K$-matrix parameterization contains $(n^2+n)$ real parameters:
 $n^2$ for $n$ copies of $\gamma_\alpha$'s and another $n$ for the $E_\alpha$'s
 while an $n$-channel Ross-Shaw parameterization has $n(n+1)/2+n$ real
 parameters, $n(n-1)/2$ parameters less than the most general
 $K$-matrix given in Eq.~(\ref{eq:K-matrix-poles}). In this paper,
 we will focus on  the two-channel case only.

\subsection{Resonance scenario in Ross-Shaw theory}
 \label{subsec:resonance_in_RossShaw}

 In this subsection, we investigate the possibility of a narrow peak
 just below the threshold of the second channel.
 In particular, this will be studied within
 the framework of two-channel Ross-Shaw theory.
 It turns out that this requirement will implement some
 constraints among the different parameters in Ross-Shaw theory.
 Later on, we will extract these parameters from our lattice data
 and therefore try to answer the question if there could exist
 a narrow peak in the elastic cross section.

 It is convenient to inspect the resonance scenario using the
 so-called $T$-matrix which is continuous across the threshold.
 Formally, it is related to the $K$-matrix via,
 \be
 K^{-1}=T^{-1}+i\;,
 \ee
 or equivalently as $T=K(1-iK)^{-1}$. The relation between the $S$-matrix
 and the $T$-matrix is given by,
 \be
 S=1+2iT\;,
 \ee
 where both $S$ and $T$ now are $2\times 2$ matrices in channel space.
 Since the scattering cross section $\sigma_{ij}$ is essentially proportional
 to $|T_{ij}|^2$, in fact we have, see Eq.~(\ref{eq:Xsections}),
 \be
 \sigma_{11}=\frac{4\pi}{k^2_1}|T_{11}|^2\;.
 \ee
 Therefore, if we write,
 \be
 T_{11}=\frac{1}{\alpha_1(E)-i}\;,
 \ee
 with $\alpha$ being real, it is seen that a resonance peak happens when $\alpha(E)=0$
 and the half-width positions are at $\alpha(E)=\pm 1$ respectively.
 Here, of course, we have neglected the energy dependence
 of the kinematic factor $1/k^2_1$ in $\sigma_{11}$ which is
 legitimate for narrow resonances.

 It is convenient to use the idea of complex phase shifts in
 two different channels.
 In this regard, one writes:
 \be
 S_{11}=e^{2i\delta^C_1}\;,
 S_{22}=e^{2i\delta^C_2}\;.
 \ee
 In order to ensure unitarity, the imaginary part of the two complex phase shifts have to
 be equal and positive:
 \be
 Im(\delta^C_1)=Im(\delta^C_2)=\varepsilon \ge 0\;.
 \ee
 The real parts of the two complex phase shifts need not have any relations.
 Comparing with the general representation in Eq.~(\ref{eq:S2-canonical}) we have
 \be
 \delta_1=Re(\delta^C_1)\;,
 \;\;\delta_2=Re(\delta^C_2)\;,\;\;
 \eta=e^{-2\varepsilon}\;,
 \ee
 and the positivity of $\varepsilon$ ensures that $\eta$ is
 a positive real number between zero and one.

 The above equations apply when the energy is above the threshold.
 Below the threshold, we have practically single-channel scattering
 and phase shift for channel one is real and that for the second
 vanishes identically: $S_{11}=e^{2i\delta_1}$, $S_{12}=S_{21}=0$, $S_{22}=1$.

 At this stage, it is important to realize a fact
 that the matrix $M$ which is equivalent to $\cot\delta$
 could develop discontinuity at the threshold. This is understandable
 since usually at the new threshold, the phase shift for the
 newly opened channel starts from zero which is a singular point
 for $\cot\delta$. However,
 the matrix $T$, which is $e^{i\delta}\sin\delta$ is perfectly
 continuous across the thresholds where $\delta=0$.
 It is therefore more convenient to use the $T$-matrix (although
 parameterized by $M$-matrix) to analyze the cross sections.

 Using the complex phase shift in channel one, we have
 \be
 T_{11}=\frac{1}{\cot\delta^C_1-i}\;,
 \;\;\cot\delta^C_1=\alpha_1(E)\;.
 \ee
 We easily obtain the following formula for $\alpha_1(E)$,
 \be
 \label{eq:general_alpha}
 \alpha_1(E)=\frac{1}{k_1(E)}\left[M_{11}(E)
 -\frac{M^2_{12}}{M_{22}(E)+\kappa_2(E)}\right]
 \;,
 \ee
 where we have substituted $-ik_2=\kappa_2$ with $\kappa_2>0$ for
 the case of a bound state in the second channel just below the threshold.
 \footnote{Thus, it is readily seen that $\alpha_1(E)$ is indeed real.}

 In the zero-range approximation, meaning that we neglect the
 effects of the effective ranges and set both $R_1$ and $R_2$ to zero,
 the elastic scattering cross section reads,
 \be
 \label{eq:sigma_e_three_parameter}
 \sigma_e=\frac{4\pi}{k^2_1
 +\left(M_{11}-\frac{M^2_{12}}{M_{22}+\kappa_2}\right)^2}
 \;,
 \ee
 where $\kappa_2=\sqrt{2m_2(E_T-E)} $ is the binding momentum in
 the second channel. The function $k^2_1$ also has a mild energy
 dependence. The resonance occurs when the second term in the
 denominator exactly vanishes, giving
 \be
 M_{11}=\frac{M^2_{12}}{M_{22}+\kappa_2}
 \;,
 \ee
 which is equivalent to,
 \be
 \kappa_2=\kappa_{2c}\equiv \frac{M^2_{12}}{M_{11}}-M_{22}
 \;.
 \ee
 For later convenience, we introduce the determinant of the $M$ matrix,
 \be
 \Delta\equiv M_{11}M_{22}-M^2_{12}\;.
 \ee
 Therefore, the value of $\kappa_2$ at which resonance occurs can be
 written as,
 \be
 \kappa_{2c}=-\frac{\Delta}{M_{11}}\;.
 \ee
 Thus, in order to have a resonance close to the threshold
 the value of $\kappa_{2c}$ has to be a positive number close to zero.
 That means the matrix $M$ has to be singular.
 This also makes sense because if the matrix $M$ is singular,
 that means $\cot\delta$ as a matrix is singular, meaning an almost
 divergent scattering length, thus signaling a large scattering
 cross section.
 On the other hand, if we were to obtain a rather large value
 for the $M$ matrix, that means the scattering phase shift matrix
 itself is small, designating a small cross section.

 It is also straightforward to work out the half-width points
 that correspond to $\alpha=\pm1$. We call them $\kappa^\pm_2$
 and they satisfy the following equation,
 \be
 \pm k_1=M_{11}-\frac{M^2_{12}}{M_{22}+\kappa^\pm_2} \;.
 \ee
 which yields,
 \be
 \kappa^\pm_2=\frac{M^2_{12}}{M_{11}\mp k_1}
 -M_{22}\;.
 \ee
 Therefore, half-width $\Gamma$ of the peak is given by,
 \be
 \Gamma=\frac{1}{2}|\kappa^+_2-\kappa^-_2|
 =\frac{k_1M^2_{12}}{|M^2_{11}-k^2_1|}\;.
 \ee

 To summarize, to characterize the narrowness of the resonance
 we would use the dimensionless ratio $R_{\mbox{narrow}}=\Gamma/k_1$ while for the
 closeness to the threshold, we would use the dimensionless
 ratio $R_{\mbox{close}}=\kappa_{2c}/k_1$. According to the above discussion,
 these two ratios read,
 \be
 R_{\mbox{close}}=-\frac{\Delta}{M_{11}k_1}\;
 \;\;
 R_{\mbox{narrow}}=\frac{M^2_{12}}{|M^2_{11}-k^2_1|}
 \;.
 \ee
 In order to have a narrow resonance just below the threshold,
 both of the above ratios have to be small.
 It is also seen that, apart from the kinematic factor $k_1$,
 all the other quantities are determined by parameters in the $M$ matrix.

 In real world, with the known information of $m_{J/\psi}=3097$MeV,
  $m_\pi=140$MeV, $m_D=1864$MeV and $m_{D^*}=2010$MeV, it is found that
  momentum $k_1=709$MeV. Therefore, for the peak of $Z_c(3900)$, taking values measured in
  Ref.:~\cite{Ablikim:2013mio}, we get,
 \ba
 R_{\mbox{close}} &=&\frac{15}{709}=0.0211\;,
 \nonumber \\
  R_{\mbox{narrow}} &=&\frac{46}{709}=0.065
  \;,
 \ea
 both of which are small numbers.
 The numbers $R_{\mbox{close}}$ and $R_{\mbox{narrow}}$ may be expressed
 in terms of the three parameters in the $M$ matrix.
 To be precise, we have,
 \be
 \label{eq:close_narrow}
 \left\{\begin{aligned}
 R_{\mbox{close}} &= \frac{M^2_{12}-M_{11}M_{22}}{M_{11}k_1}\;,
 \\
 R_{\mbox{narrow}} &=\frac{M^2_{12}}{M^2_{11}-k^2_1}\;.
 \end{aligned}\right.
 \ee
 Therefore,   given the parameters $M_{22}$, $R_{\mbox{close}}$ and $R_{\mbox{narrow}}$
 we can get the values of $M_{11}$ and $M_{12}$
 which can then be compared with what we obtain from the lattice data.
 We will call these conditions the closeness and narrowness condition in the following.

 However, since we are not having a physical pion mass in the simulation,
 we need to relax this constraint. For a generic pion mass, we have,
 \be
 k^2_{10}=\left[\frac{(m_D+m_{D^*})^2+m^2_\pi-m^2_{J/\psi}}{2(m_D+m_{D^*})}\right]^2-m^2_\pi
 \;.
 \ee
 In our simulation, the mass of the relevant mesons are given in the following table
 where all expressed in lattice units:
 \ba
 m_\pi &=&0.1416(1),\;
 m_{J/\psi}=1.2985(3),\;
 \nonumber \\
 m_{D^*} &=& 0.8875(12),\;
 m_D=0.7967(4)\;,
 \ea
 Substitute into the expression for $k_{10}$, we get,
 \be
 k_{10}=0.3174\;.
 \ee
 in lattice units, corresponding to about $720$MeV in physical unit,
 which is close to real world value of $709$MeV.
 Thus, we may demand that both ratios are close to its real world values in our analysis.
 Note that the matrix elements in the $M$-matrix also has dimension of momentum.
 Therefore, it is convenient to measure everything in units of $k_{10}$.
 Within this unit system, the closeness and narrowness conditions read,
 \be
 \label{eq:close_narrow_k1unit}
 \left\{\begin{aligned}
 R_{\mbox{close}} &= \frac{M^2_{12}-M_{11}M_{22}}{M_{11}}\;,
 \\
 R_{\mbox{narrow}} &=\frac{M^2_{12}}{M^2_{11}-1}\;.
 \end{aligned}\right.
 \ee
 where everything is measured in units of $k_{10}$. This is more convenient when
 we analyze our data later on in subsection~\ref{subsec:confidence_level}.

 \section{One- and two-particle operators and correlators}
 \label{sec:operators}

 Single-particle and two-particle energies are measured
 in Monte Carlo simulations by measuring corresponding correlation functions,
 which are constructed from appropriate interpolating operators with definite symmetries.
 For the $D\bar{D}^*$ channel, we basically adopted the same set of operators in our previous study, see
 Sec. III in Ref.~\cite{Chen:2014afa}. Below, we will take this channel as an example.
 Operators in other channels can also be constructed accordingly.

 \subsection{One- and two-particle operators and correlators}

 Let us list the interpolating operators for the charmed mesons,  we utilize the following local interpolating
 fields in real space for $D$ mesons:
 \be
 \label{eq:single_operators_defs}
 [D^+]:\ \calP^{(d)}(\bx,t) = [\bar{d}\gamma_5 c](\bx,t)\;,
 \ee
 together with the interpolating operator for its anti-particle ($D^-$):
 $\bar{\calP}^{(d)}(\bx,t)=[\bar{c}\gamma_5d](\bx,t)=[\calP^{(d)}(\bx,t)]^\dagger$.
 In the above equation, we have also indicated the quark flavor content of the operator
 in front of the definition inside the square bracket. So, for example, the operator in Eq.~(\ref{eq:single_operators_defs})
 will create a $D^+$ meson when acting on the QCD vacuum.
 Similarly, one defines $\calP^{(u)}$ and $\bar{\calP}^{(u)}$ with the
 quark fields $d(\bx,t)$ in Eq.~(\ref{eq:single_operators_defs}) replaced by $u(\bx,t)$.
 In an analogous manner, a set of operators $\calV^{(u/d)}_i$ are constructed
 for the vector charmed mesons $D^{*\pm}$ with the $\gamma_5$ in $\calP^{(u/d)}$ replaced
 by $\gamma_i$.
 A single-particle state with definite three-momentum $\bk$ is
 defined accordingly via Fourier transform, e.g.:
 \be
 \label{eq:singlemeson_untwist}
  \calP^{(u/d)}(\bk,t)=\sum_\bx \calP^{(u/d)}(\bx,t)e^{-i \bk \cdot \bx}.
 \ee
 The conjugate of the above operator is:
 \be
  [\calP^{(u/d)}(\bk,t)]^\dagger=\sum_\bx [\calP^{(u/d)}(\bx,t)]^\dagger
  e^{+i \bk \cdot \bx}\equiv\bar{\calP}^{(u/d)}(-\bk,t).
 \ee
 Similar relations also hold for $\calV^{(u/d)}_i$ and $\bar{\calV}^{(u/d)}_i$.
 The interpolating operators for $J/\psi$, $\pi$, $\rho$ and $\eta_c$ are formed accordingly.

 To form the two-particle operators, one has to consider the corresponding
 internal quantum numbers.
 Since our target state $Z^\pm_c(3900)$ state likely carries $I^G(J^{PC})=1^+(1^{+-})$,
 we use:
 \be
 1^+(1^{+-}):\ \left\{\begin{aligned}
 & D^{*+}\bar{D}^0+\bar{D}^{*0}D^+
 \\
 & D^{*-}\bar{D}^0+\bar{D}^{*0}D^-
 \\
 & [D^{*0}\bar{D}^0-D^{*+}D^-]
 + [\bar{D}^{*0}D^0-D^{*-}D^+]
 \end{aligned} \right.
 \ee

 Therefore, in terms of the operators
 defined in Eq.~(\ref{eq:single_operators_defs}), we have used
 \be
 \calV^{(d)}_i(\bk,t)\bar{\calP}^{(u)}(-\bk,t)
 +\bar{\calV}^{(u)}_i(\bk,t)\calP^{(d)}(-\bk,t)
 \;,
 \ee
 for a pair of charmed mesons with back-to-back momentum $\bk$.

 On the lattice, the rotational symmetry group $SO(3)$ is broken down
 to the octahedral group $O(Z)$.
 We use the following operator to create the two charmed meson state from the vacuum,
 \begin{widetext}
 \be
 \label{eq:two-particle-operator-nontwist}
 \calO^i_\alpha(t)=\sum_{R\in O(Z)}
 \left[\calV^{(d)}_i(R\circ\bk_\alpha,t)\bar{\calP}^{(u)}(-R\circ\bk_\alpha,t)
 +\bar{\calV}^{(u)}_i(R\circ\bk_\alpha,t)\calP^{(d)}(-R\circ\bk_\alpha,t)
 \right]
 \;,
 \ee
 \end{widetext}
 where $\bk_\alpha$ is a chosen three-momentum mode. The index
 $\alpha=1,\cdots,N$ with $N$ being the number of momentum modes
 considered in a corresponding channel. In this calculation, we have
 chosen $N=4$ for all channels with $\bk^2_\alpha=(2\pi/L)^2\bn^2_\alpha$,
  $\bn^2_\alpha=0,1,2,3$.
 In the above equation,  the notation $R\circ\bk_\alpha$
 represents the momentum obtained from $\bk_\alpha$ by applying the operation $R$ on $\bk_\alpha$.

 In an analogous fashion, single and two-particle operators are constructed in other channels.
 For example, for the $J/\psi\pi$ channel, one simply replaces the operators
 for $D^*$ and $D$ by the corresponding ones for $J/\psi$ and $\pi$, respectively.

 \subsection{Correlation functions}

 One-particle correlation function, with a definite three-momentum $\bk$,
 for the vector and pseudo-scalar charmed mesons are defined respectively as,
 \ba
 \label{eq:single-particle-correlators}
  C^\calV(t,\bk) &=& \langle \calV^{(u/d)}_i(\bk,t)\bar{\calV}^{(u/d)}_i(-\bk,0)\rangle\;,
  \nonumber \\
  C^\calP(t,\bk) &=& \langle \calP^{(u/d)}(\bk,t)\bar{\calP}^{(u/d)}(-\bk,0)\rangle
  \;.
 \ea
 From these correlation functions, including similar ones for other particles
 discussed in this study, it is straightforward to obtain the single
 particle energies for various lattice momenta $\bk$, enabling us to
 check the dispersion relations for all single particles.

 We now turn to more complicated two-particle correlation
 functions. Generally speaking, we need to evaluate a (hermitian) correlation
 matrix of the form:
 \be
 \label{eq:correlation}
  C_{\alpha\beta}(t)=\langle \calO^{i\dagger}_{\alpha}(t) \calO^i_{\beta}(0) \rangle,
 \ee
 where $\calO^i_\alpha(t)$ represents the two-particle operator defined
 in Eq.~(\ref{eq:two-particle-operator-nontwist}) in a particular channel.
  Two particle energies that are to
 be substituted into L\"uscher's formula are obtained from this correlation
 matrix by solving the so-called generalized eigenvalue problem (GEVP):
 \footnote{We have used the matrix notation.}
 \be
 \label{eq:GEVP}
 C(t)\cdot v_\alpha(t,t_0)=\lambda_\alpha(t,t_0)C(t_0)\cdot v_\alpha(t,t_0)\;,
 \ee
 with $\alpha=1,2,\cdots,N_{op}$ and $t>t_0$.
 The eigenvalues $\lambda_\alpha(t,t_0)$ can be shown to behave like
 \be
 \label{eq:exponential}
 \lambda_\alpha(t,t_0)\simeq e^{-E_\alpha(t-t_0)}+\cdots\;,
 \ee
 where $E_\alpha$ being the eigenvalue of the Hamiltonian for the system
 which is the quantity to be substituted into L\"uscher's formula to extract
 the scattering information.
 The parameter $t_0$ is tunable and one could optimize the calculation
 by choosing $t_0$ such that the correlation function is more or less dominated
 by the desired eigenvalues at that particular $t_0$ (preferring a larger $t_0$) with
 an acceptable signal to noise ratio (preferring a smaller $t_0$).
 The eigenvectors $v_\alpha(t,t_0)$ are orthonormal with respect to the
 metric $C(t_0)$, $v^\dagger_\alpha C(t_0) v_\beta=\delta_{\alpha\beta}$ and
 they contain the information of the overlaps of the original operators with
 the eigenvectors.


 \subsection{sLapH smearing}

 To enhance the signal for the correlation matrix functions defined in the previous subsection,
 we have utilized the  stochastic  Laplacian  Heavyside  smearing (sLapH smearing) as discussed in Ref.~\cite{Morningstar:2011ka}.
 Perambulators for the charm and light quarks are obtained using diluted stochastic sources.
 We adopted the same dilution procedure as in Ref.~\cite{Helmes:2015gla}, see Sec. 2.1 in that reference
 for further details. The correlation functions are then constructed from these perambulators.

 \subsection{Singling out the most relevant two channels}

 \begin{figure*}[htb]
  {\resizebox{0.9\textwidth}{!}{\includegraphics{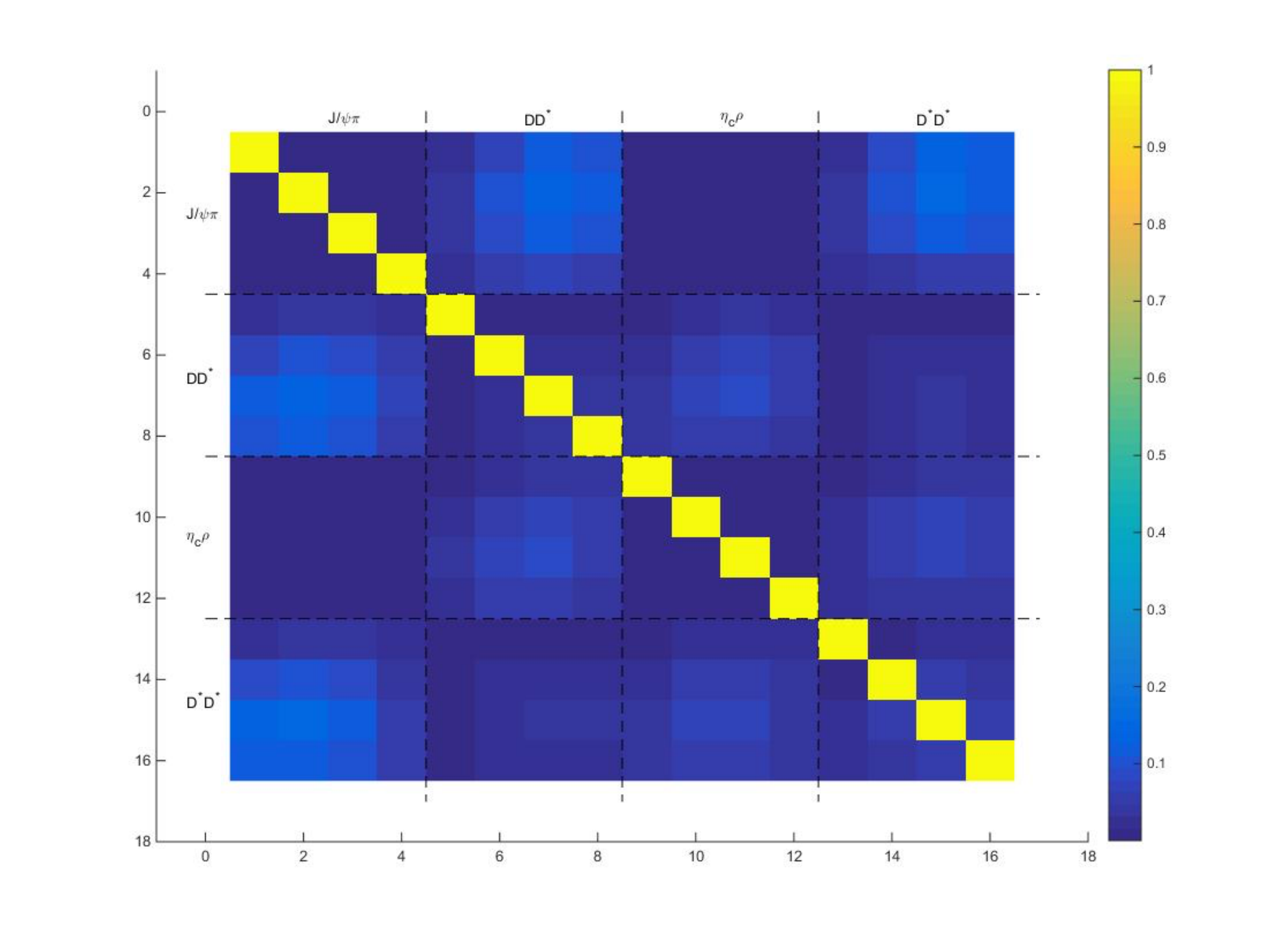}}}
 \caption{Density plot for the magnitude of the correlation matrix $\tilde{C}_{\alpha\beta}(t)$ as defined
 in Eq.~(\ref{eq:cross_correlation}) at $t=10$  obtained from  the ensemble. Four channels have been included, each with 4 different back-to-back momentum. It is seen that the coupling between $J/\psi\pi$ and
 $D\bar{D}^*$ is the most relevant one. \label{fig:density}}%
  \end{figure*}
 There exist four relevant channels in the energy regime we are investigating, namely $J/\psi\pi$, $D\bar{D}^*$, $\eta_c\rho$ and $D^*\bar{D}^*$, with increasing thresholds. It is suggested by HALQCD collaboration that the lowest three channels had strong couplings among each other~\cite{Ikeda:2016zwx,Ikeda:2017mee} .
 To verify this, we took all four channels and construct the cross-correlation matrix $\tilde{C}(t)$ as defined in Eq.~(\ref{eq:cross_correlation}).
 Within each of the four channels, in the center of mass frame,
 we construct the two meson operators with back-to-back three-momentum
 characterized by a three-dimensional integer $\bn$ as in Eq.~(\ref{eq:free_k}),
 starting from $\bn^2=0$ to $\bn^2=3$. Then, the correlation matrix is measured using
 the stochastically estimated perambulators obtained from the ensemble listed in Table~\ref{tab:parameter}.
 By inspecting the magnitude of the off-diagonal matrix elements relative to the diagonal ones
 we get a feeling about the coupling among these channels.

 To visualize this, starting from the correlation matrix $C_{\alpha\beta}(t)$ defined
 in Eq.~(\ref{eq:correlation}), we define a new cross-correlation matrix $\tilde{C}(t)$
 at a particular temporal separation $t$ as:
 \be
 \label{eq:cross_correlation}
 \tilde{C}_{\alpha\beta}(t)=C_{\alpha\beta}(t)/\sqrt{|C_{\alpha\alpha}(t)C_{\beta\beta}(t)|}
 \;,
 \ee
 such that the diagonal matrix elements of $\tilde{C}(t)$ are normalized to unity.
 Since there are four channels and each channel has four different momenta, the
 matrix $\tilde{C}(t)$ is a $16\times 16$ matrix.
 If there were no cross-channel couplings, the matrix would be block diagonal within each channel.

 The magnitude of $\tilde{C}$ is shown in Fig.~\ref{fig:density} for $t=10$ for the case of
 four channels, namely $J/\psi\pi$, $D\bar{D}^*$, $\eta_c\rho$ and $D^*\bar{D}^*$.
 The column of the matrix is numbered from left to right according to the channels:
 $J/\psi\pi$, $D\bar{D}^*$, $\eta_c\rho$ and $D^*\bar{D}^*$. Within each channel,
 it is numbered according to increasing $\bn^2$ for the back-to-back momenta $\bk$.
 Similar numbering is adopted for the row of the matrix, from top to bottom.
 Thus the $16\times 16$ matrix is made up of $4\times 4$ block matrices, with
 each block of $4\times 4$ matrix representing the scattering within a particular
 single channel. It is seen from the figure that the coupling among channels do exist, with the $D\bar{D}^*$ and $J/\psi\pi$ being the most prominent ones.
 We remark that, this quantity $\tilde{C}$ by itself is not a physical quantity. In fact it is also dependent
 on the time separation $t$. Since $J/\psi\pi$ being the lightest channel among the four,
 its mixture with other channels will get magnified relative to other channels as $t$ increases.
 Nevertheless, the magnitude of the off-diagonal matrix elements of $\tilde{C}$
 still offers us a qualitative description for the coupling among different channels.
 Since the number of parameters increases quadratically with the number of channels,
 in this study we are aiming at only two coupled channels.
 Therefore, in the following,
 we will focus on the two channels that are coupled most strongly, namely $D\bar{D}^*$ and $J/\psi\pi$.
 Not surprisingly, these two channels that are coupled most strongly as indicated by our simulation also
 coincides with what the experimental data suggested,  see e.g. Ref.~\cite{Ablikim:2013xfr}
 where the channel $D\bar{D}^*$ is found to be the dominant decay channel for $Z_c(3900)$
 while the channel $J/\psi\pi$ is the discovery channel for $Z_c(3900)$.

 \section{Simulation details and results}
 \label{sec:simulation_details}

 In this paper, we have utilized $N_f=2+1+1$ twisted mass gauge field configurations
 generated by European Twisted Mass Collaboration (ETMC) at $\beta=1.9$ corresponding
 to a lattice spacing $a=0.0863$fm in physical units.
 Details of the relevant parameters are summarized in table~\ref{tab:parameter}.
\begin{table}
\centering \caption{Simulation parameters in this study. The lattice spacing is
about $0.0863$fm in physical unit while the pion mass is roughly $320$MeV. \label{tab:parameter}}
\begin{tabular}{c|c|c|c|c|c|c}
\hline
 ensemble & $\beta$ & $a\mu_l$ & $a\mu_\sigma$ & $a\mu_\delta$ & $(L/a)^3\times T/a$ & $N_{conf}$ \\
\hline \hline
A40.32 & $1.9$ & $0.0040$ & $0.150$ & $0.190$ & $32^3\times 64$  & 250\\
\hline
\end{tabular}
\end{table}

 For the measurements related to the charm sector,
 we have used the Osterwalder-Seiler type action with a
 fictitious $c'$ quark introduced that is degenerate with $c$~\cite{Frezzotti:2004wz}.
 They form an $SU(2)$ doublet characterized by a quark mass parameter $\mu_c$.
 The up and down quark mass values are also degenerate and the parameter $\mu_l$ is
 fixed to the same value as in the sea corresponding to pion mass $m_\pi=320$MeV.
 For the charm quark, the parameter $\mu_c$ is fixed in
 such a way that the mass of $J/\psi$ on
 the lattice corresponds to about $2.96$GeV.

 \subsection{Single meson correlators and dispersion relations}

 \begin{figure*}[htb]
 {\resizebox{0.48\textwidth}{!}{\includegraphics{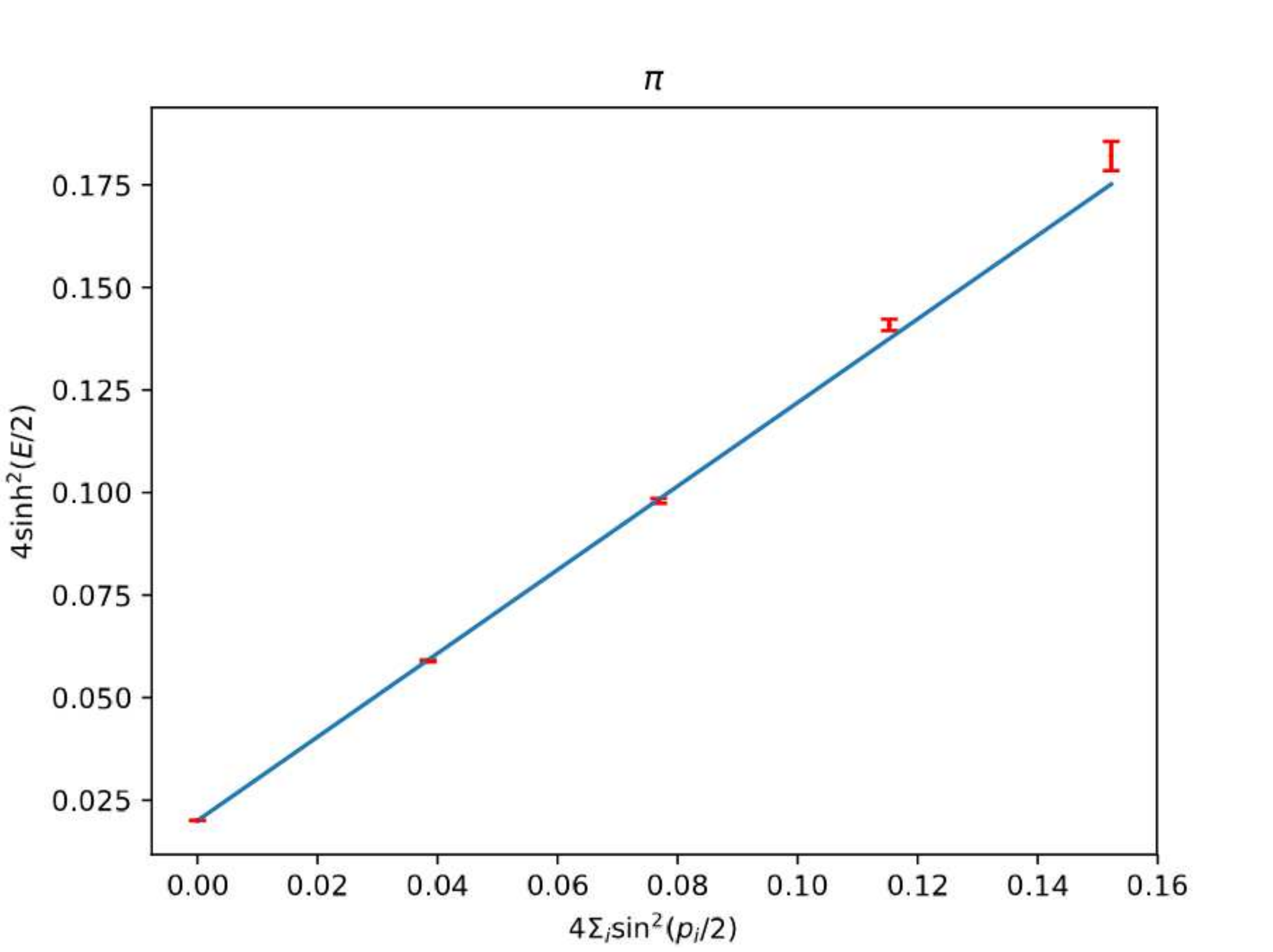}}}
  {\resizebox{0.48\textwidth}{!}{\includegraphics{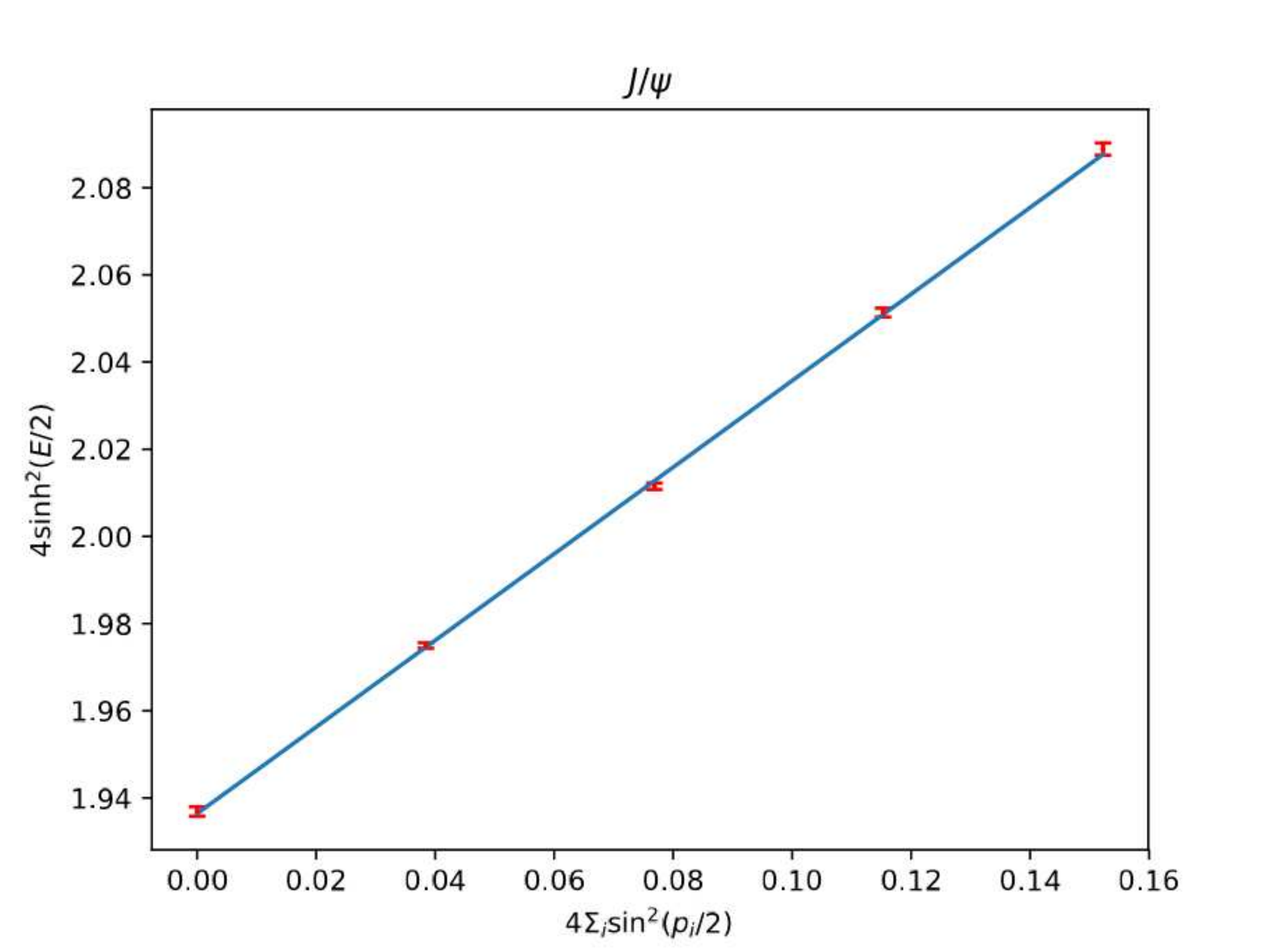}}}
  {\resizebox{0.48\textwidth}{!}{\includegraphics{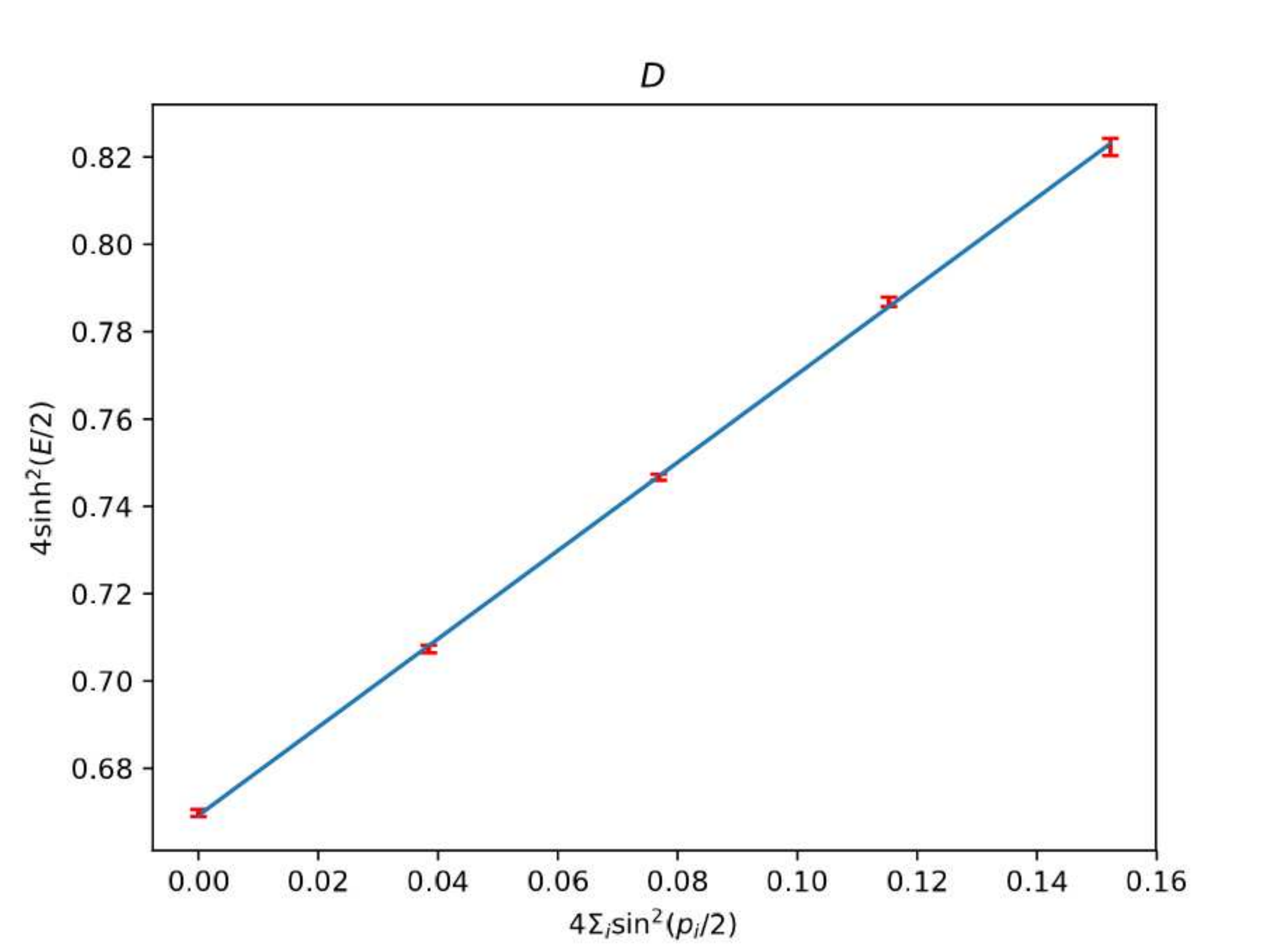}}}
  {\resizebox{0.48\textwidth}{!}{\includegraphics{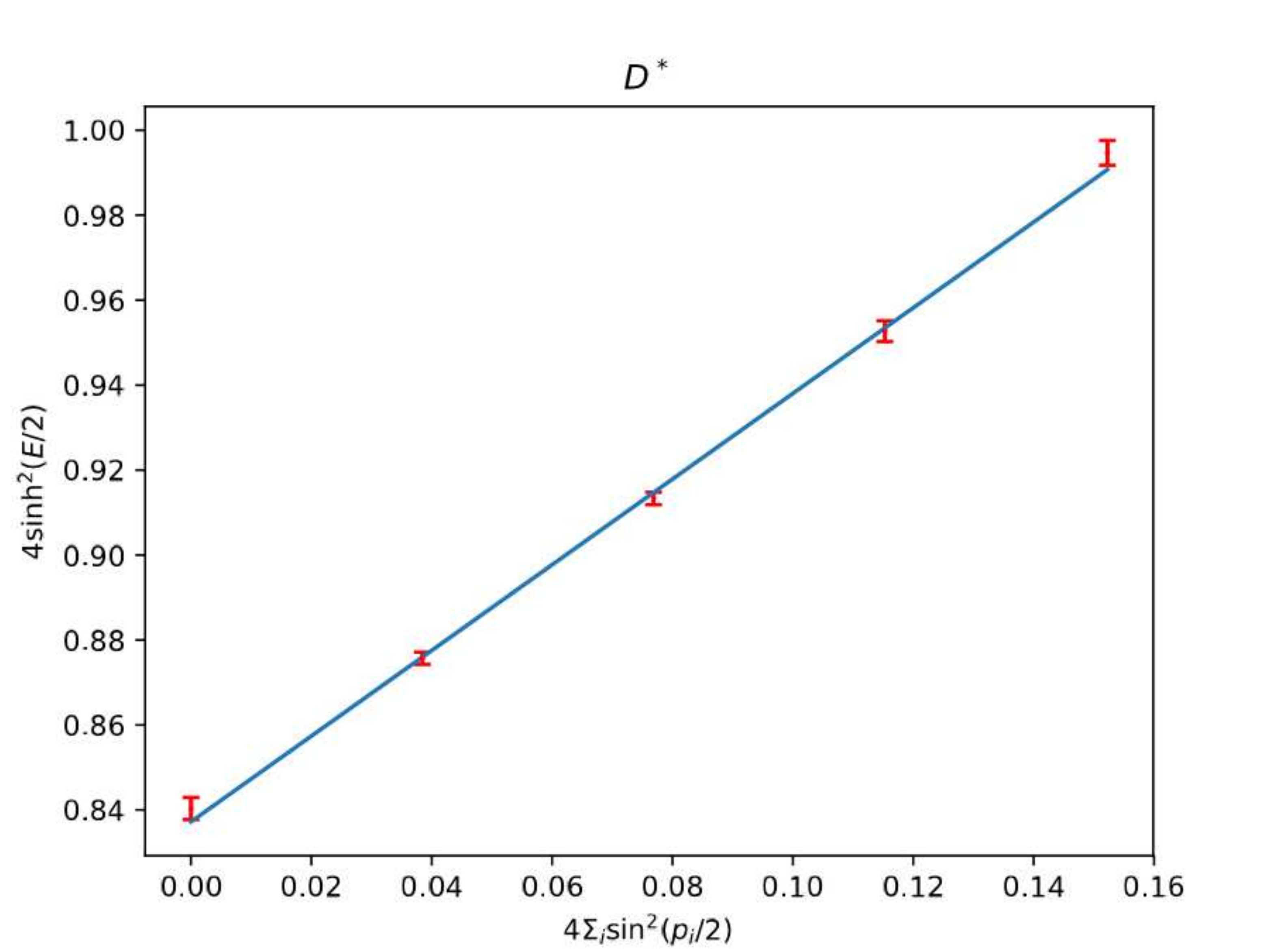}}}
 \caption{Single meson dispersion relations (the lattice version) for $J/\psi$, $\pi$, $D$ and $D^*$, respectively.
 Data points are obtained from corresponding single meson correlators and the lines are linear fits using Eq.~(\ref{eq:dispersion}).
 \label{fig:dispersion}}%
  \end{figure*}
 We first check the single meson dispersion relations
 for the mesons involved: $\pi$, $J/\psi$, $D$ and $D^*$.
 This is crucial since we need to have a well established single
 meson states within a finite box in order to utilize L\"uscher's formalism.
 We have attempted to fit the single meson correlators with various momenta
 using both the continuum and the lattice version of the dispersion relations:
 \be
 \label{eq:dispersion}
 \begin{aligned}
 E^2(\bp) &=m^2+Z_{\mbox{cont.}}\bp^2\;,
 \\
 4\sinh^2\left(\frac{E(\bp)}{2}\right)&=
 4\sinh^2\left(\frac{m}{2}\right)+Z_{\mbox{latt.}}
 \sum^3_{i=1}4\sin^2\left(\frac{p_i}{2}\right)\;.
 \end{aligned}
 \ee
 It turns out that both works fine  for small enough $\bp^2$
 except that the lattice ones are better in the sense that the slope
 $Z_{\mbox{latt.}}$ is closer to unity than the values for $Z_{\mbox{cont.}}$.
 In Fig.~\ref{fig:dispersion}, the lattice version of these dispersion relations are illustrated
 for the above listed mesons. Data points are obtained from the corresponding
 single-meson correlators while the lines representing the linear fits
 using Eq.~(\ref{eq:dispersion})

 \subsection{The two-particle energy levels}

 \begin{figure}[htb]
  {\resizebox{0.4\textwidth}{!}{\includegraphics{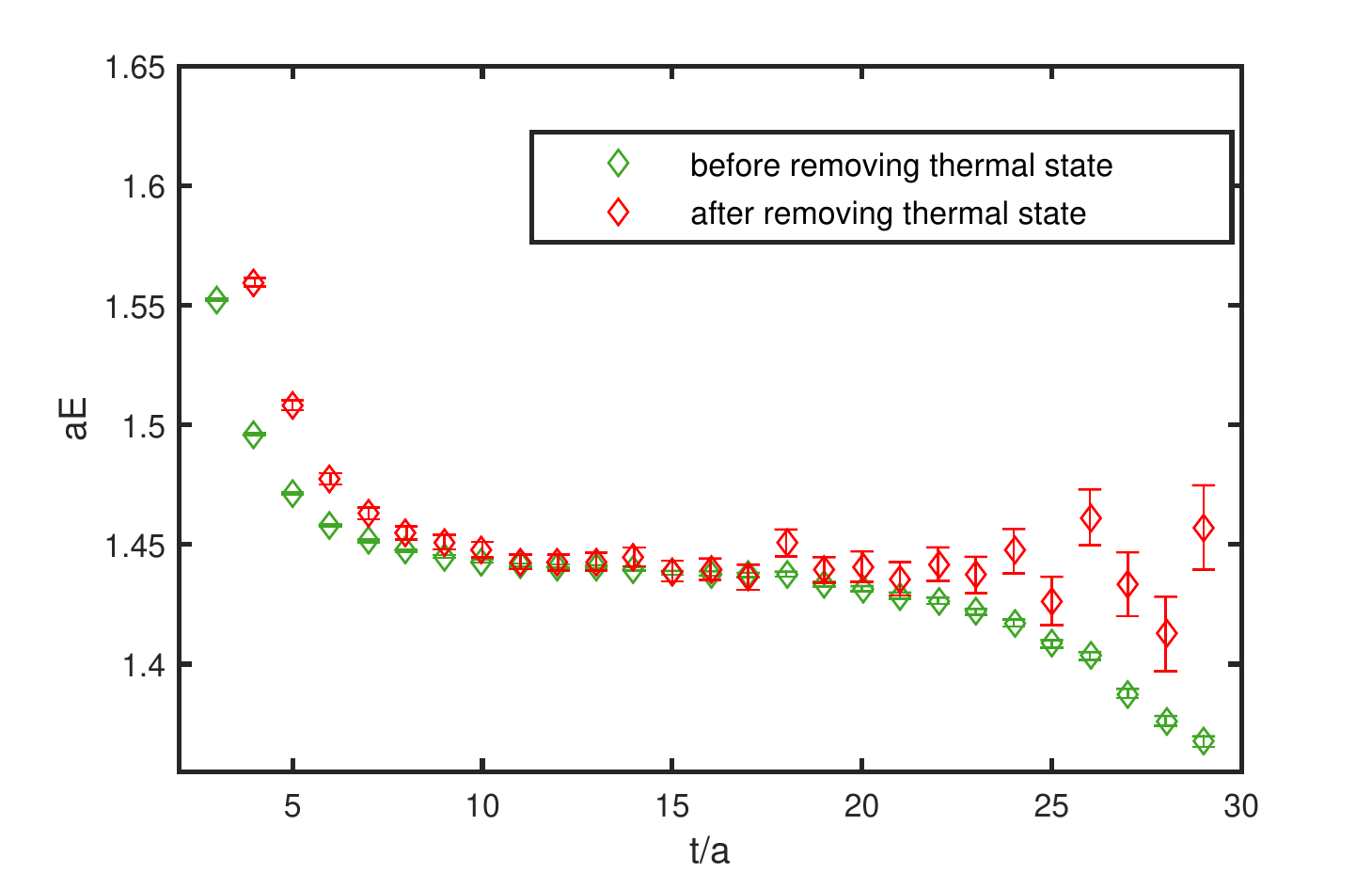}}}
 \caption{Effective mass plateaus for one of our energy eigenvalues both
 with and without the thermal state removal. \label{fig:removal}}%
  \end{figure}

\begin{table}
\centering \caption{Energy eigenvalues obtained in the coupled two channel scattering.  \label{tab:deltaE}}
\begin{tabular}{c|c|l|l}
\hline
 Eigenvalues & $[t_{\min},t_{\max}]$ & $E_\alpha$  & $\chi^2/N_{d.o.f}$ \\
\hline
 $\lambda_0$ & $[14,24]$ & $1.4411(7)$ & $0.6$\\
\hline
 $\lambda_1$ & $[11,24]$ & $1.5547(6)$ & $1.1$\\
\hline
 $\lambda_2$ & $[10,22]$ & $1.6375(10)$ & $0.7$\\
\hline
 $\lambda_3$ & $[12,22]$ & $1.6868(23)$ & $0.6$\\
\hline
$\lambda_4$ & $[11,22]$ & $1.7058(31)$ & $0.3$\\
\hline
$\lambda_5$ & $[12,18]$ & $1.7319(33)$ & $0.4$\\
\hline
$\lambda_6$ & $[12,18]$ & $1.7610(63)$ & $0.4$\\
\hline
$\lambda_7$ & $[9,14]$ & $1.8152(30)$ & $0.6$\\
\hline\hline
\end{tabular}
\end{table}
 To extract the two-particle energy eigenvalues, we adopt
 the usual L\"uscher-Wolff method~\cite{luscher90:finite}.
 For this purpose,  a new matrix $\Omega(t,t_0)$ is defined as:
 \begin{eqnarray}
  \Omega(t,t_0)=C(t_0)^{-{1\over2}}C(t)C(t_0)^{-{1\over 2}},
 \end{eqnarray}
 where $t_0$ is a reference time-slice. Normally one picks a
 $t_0$ such that the signal is good and stable.
 The energy eigenvalues for the two-particle system are
 then obtained by diagonalizing the matrix $\Omega(t,t_0)$.
 The eigenvalues of the matrix has the usual exponential decay behavior
 as described by Eq.~(\ref{eq:exponential}) and therefore
 the exact energy $E_\alpha$ can be extracted from
 the effective mass plateau of the eigenvalue $\lambda_\alpha$.

 Since we are only interested in the the two-channel scenario,
 we focus on the correlation sub-matrix in $J\psi\pi$ and $D\bar{D}^*$ sector.
 The correlation matrix is therefore $8\times 8$ and the GEVP process yields $8$ energy levels.
 In order to be able to obtain a good plateau for each of these energy levels,
 we have tried to remove the so-called thermal-state contaminations arising from
 the lightest $J/\psi\pi$ state, see e.g. Ref.~\cite{Helmes:2015gla,Dudek:2012gj}.
 In Fig.~\ref{fig:removal} we show a typical behavior for one of the energy eigenvalue plateau plot
 both with and without this procedure. It is seen that, after performing
 the so-called thermal state removal procedure, the plateau behavior is greatly improved.

 \begin{figure*}[htb]
  {\resizebox{0.24\textwidth}{!}{\includegraphics{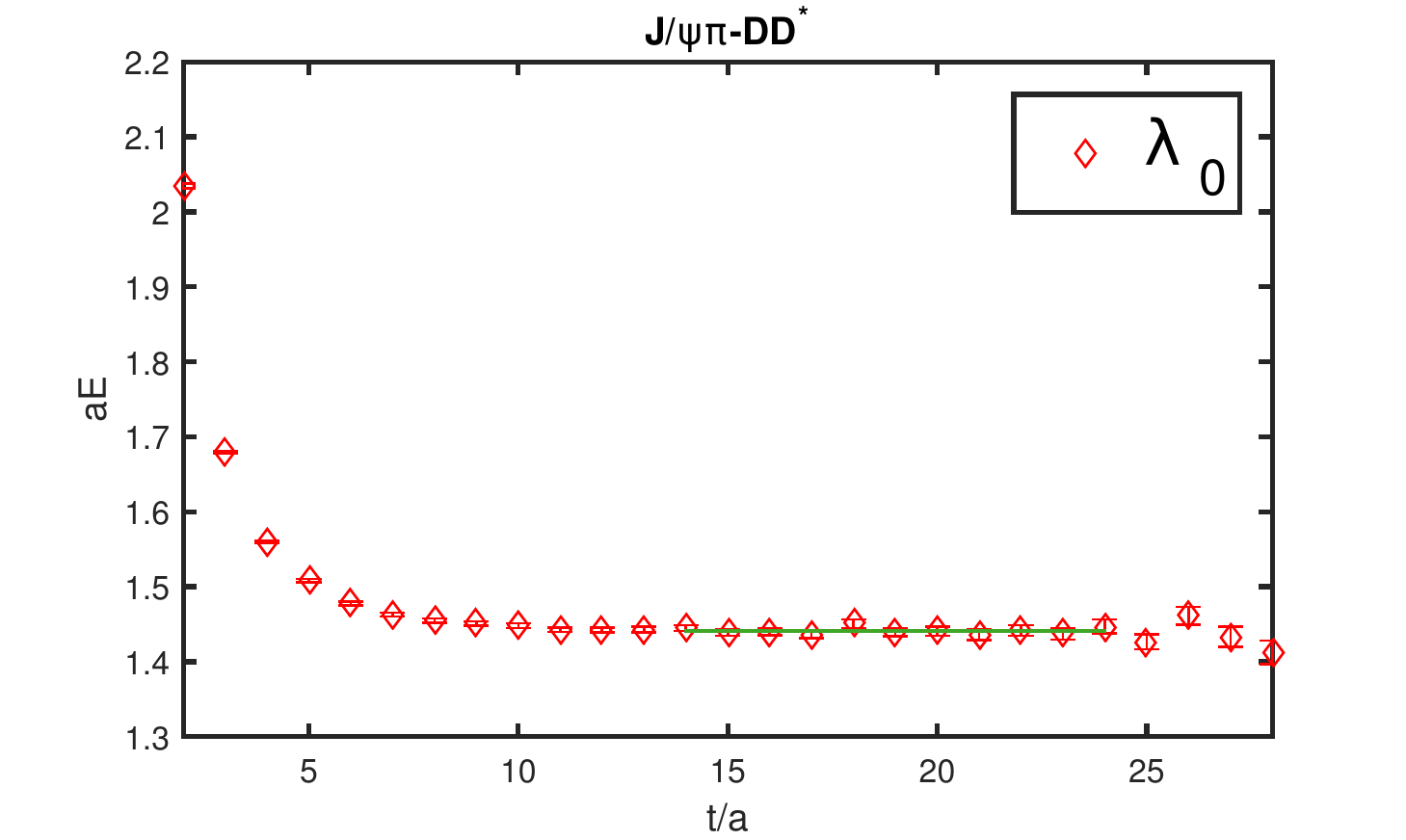}}}
{\resizebox{0.24\textwidth}{!}{\includegraphics{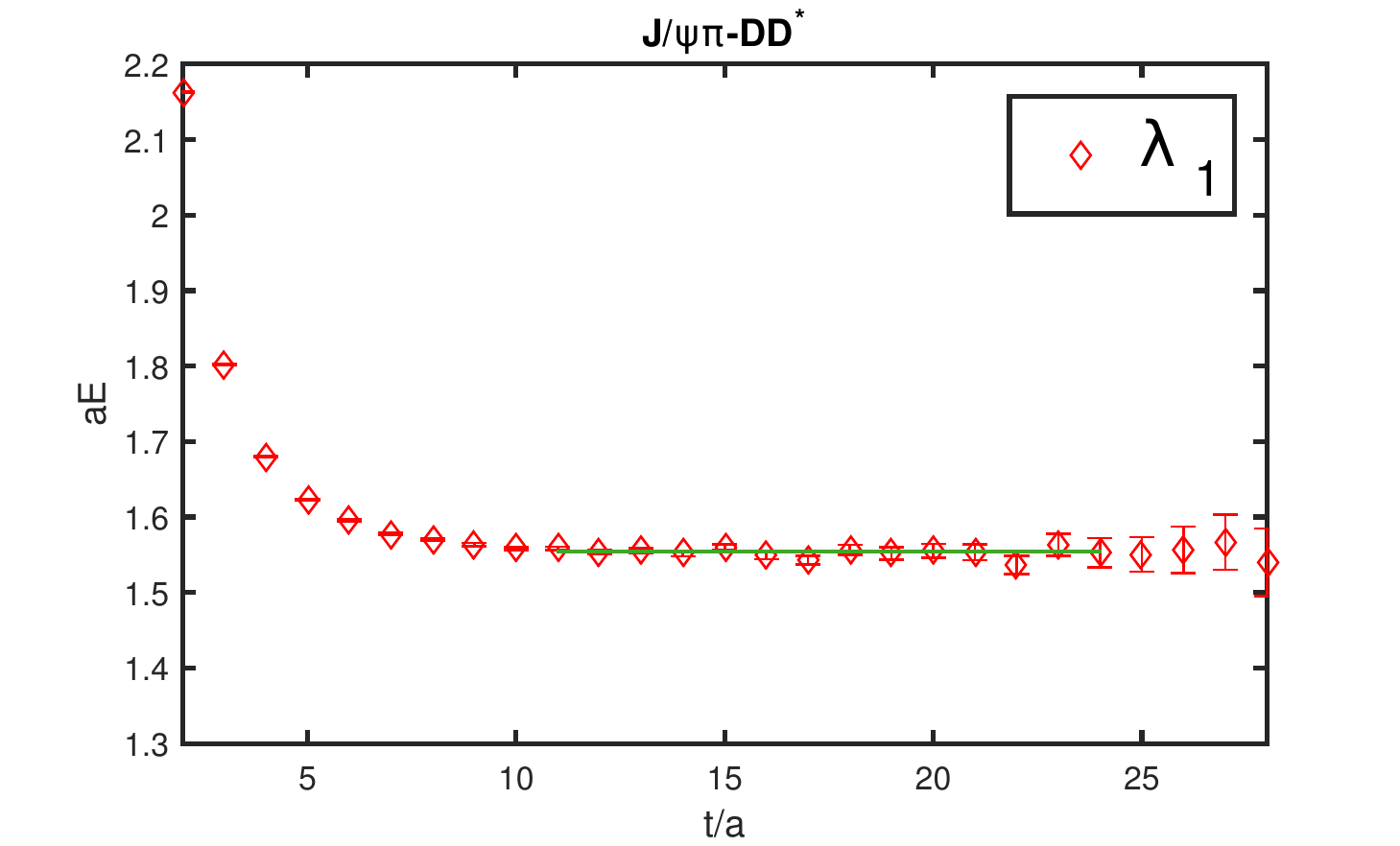}}}
{\resizebox{0.24\textwidth}{!}{\includegraphics{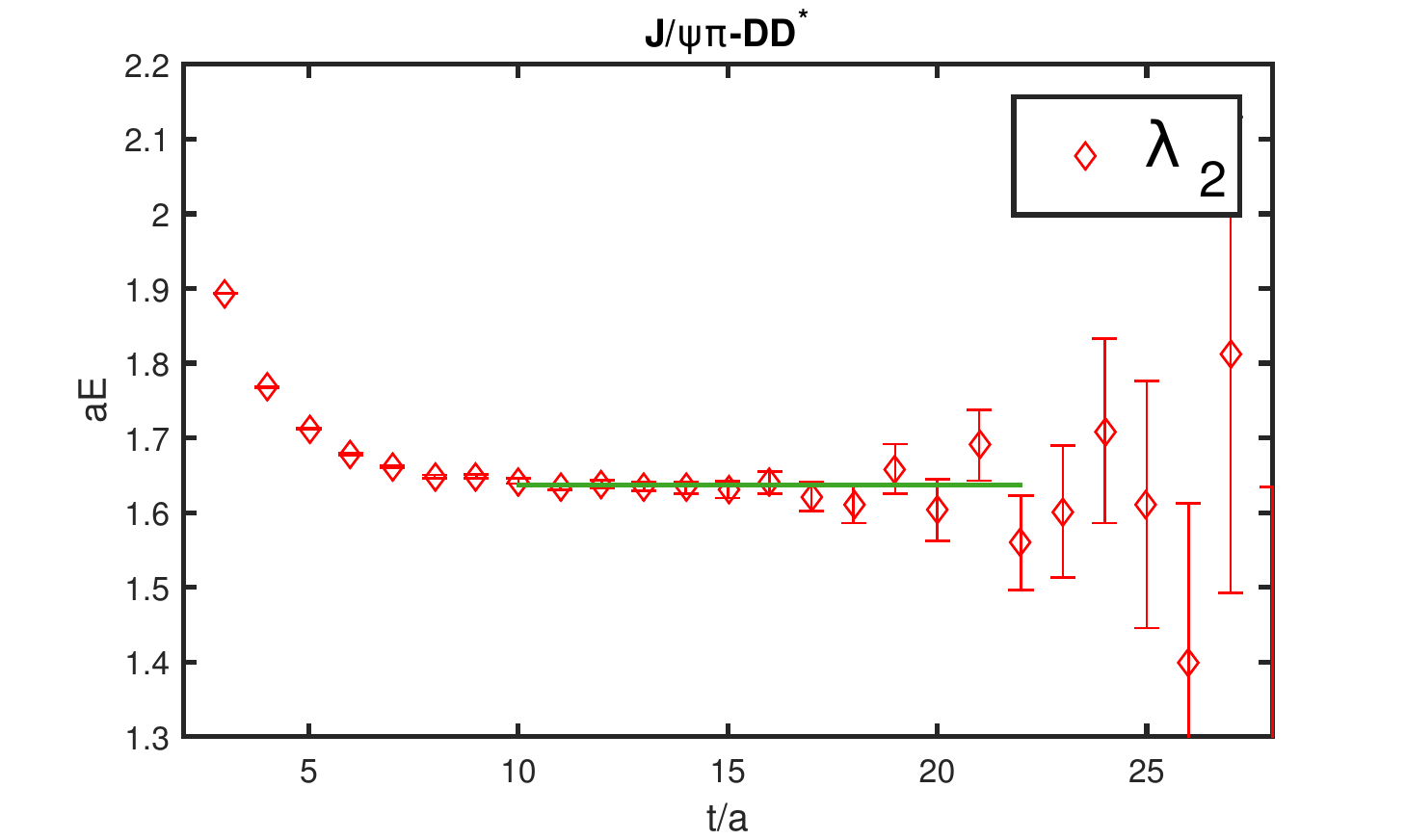}}}
{\resizebox{0.24\textwidth}{!}{\includegraphics{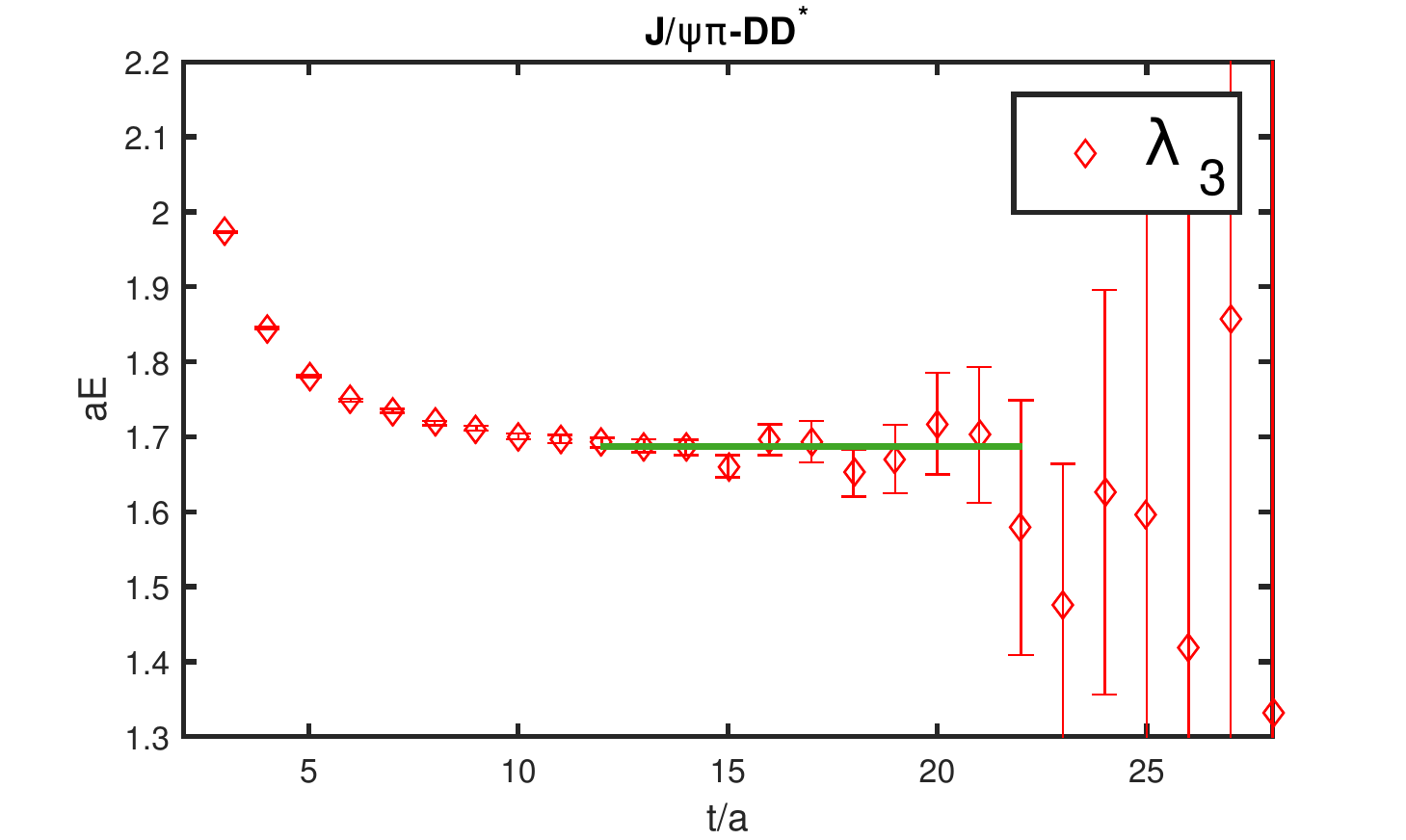}}}
{\resizebox{0.24\textwidth}{!}{\includegraphics{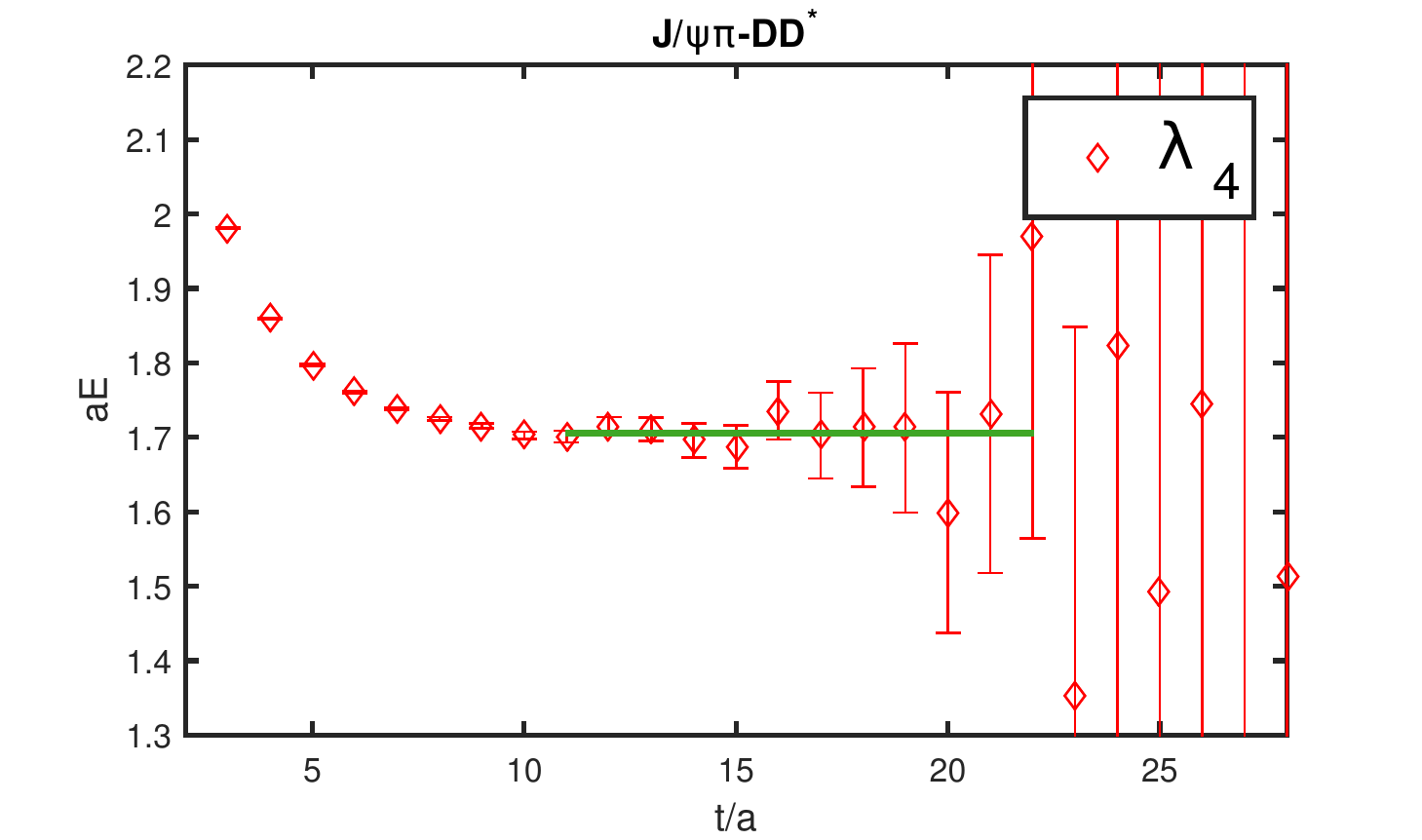}}}
{\resizebox{0.24\textwidth}{!}{\includegraphics{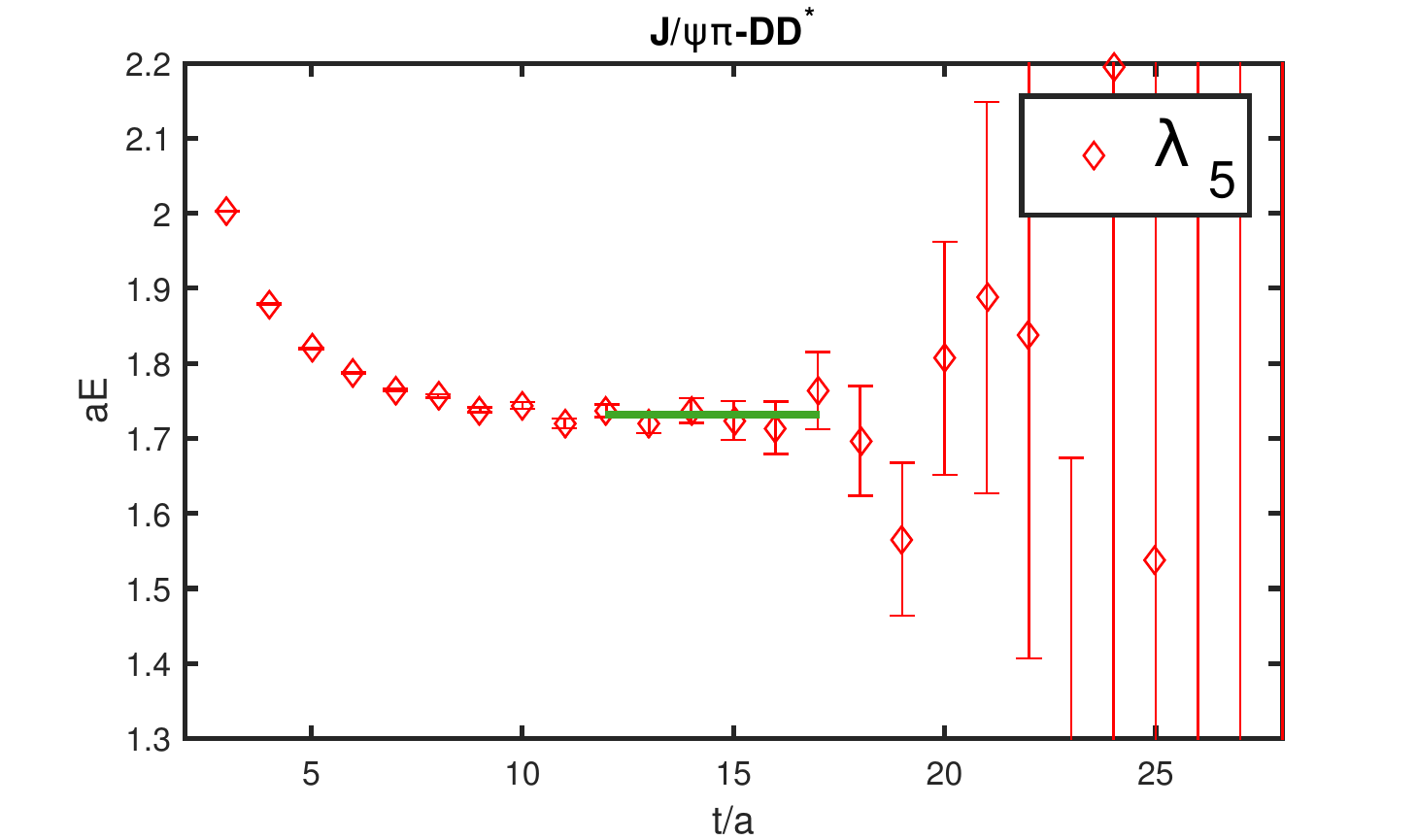}}}
{\resizebox{0.24\textwidth}{!}{\includegraphics{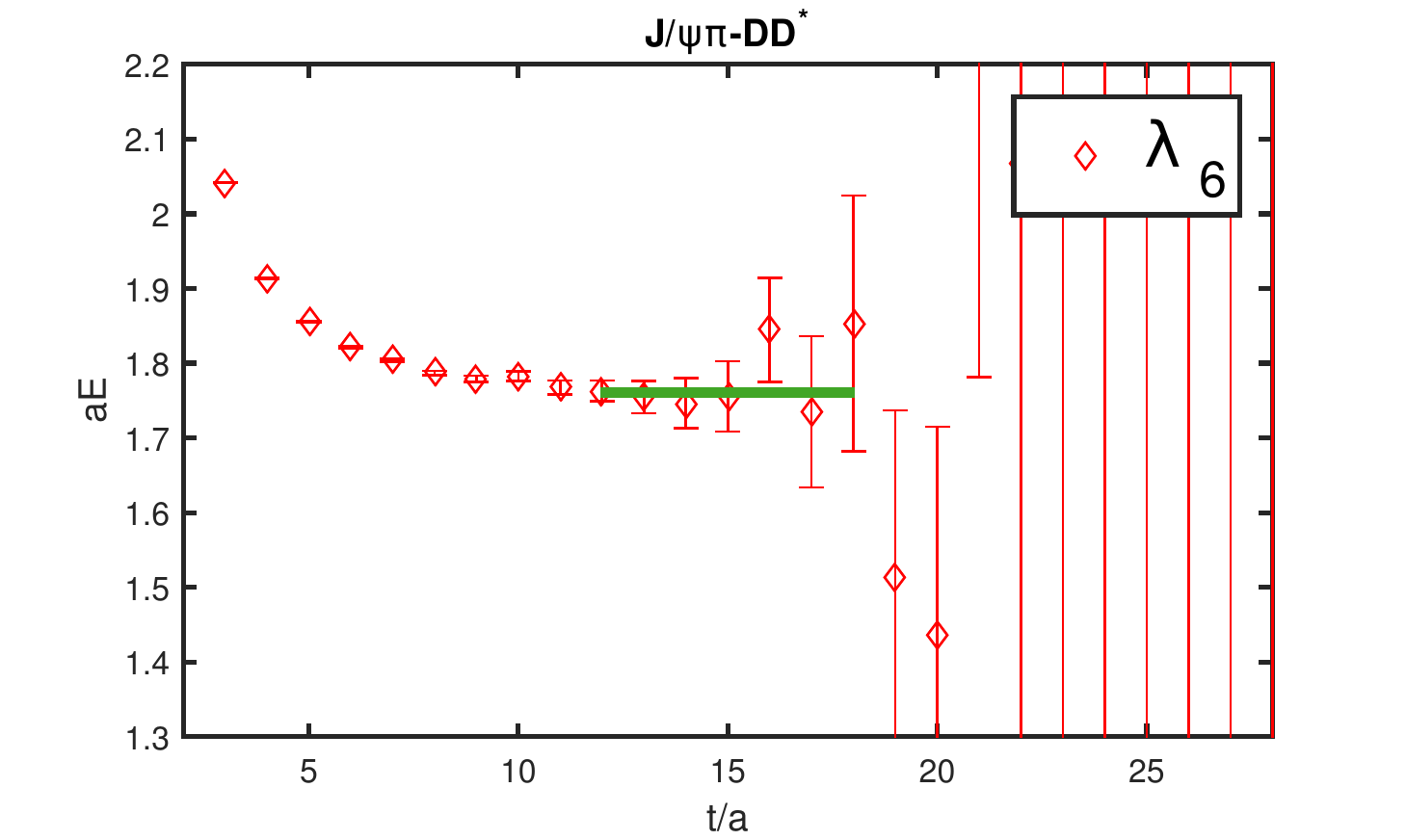}}}
{\resizebox{0.24\textwidth}{!}{\includegraphics{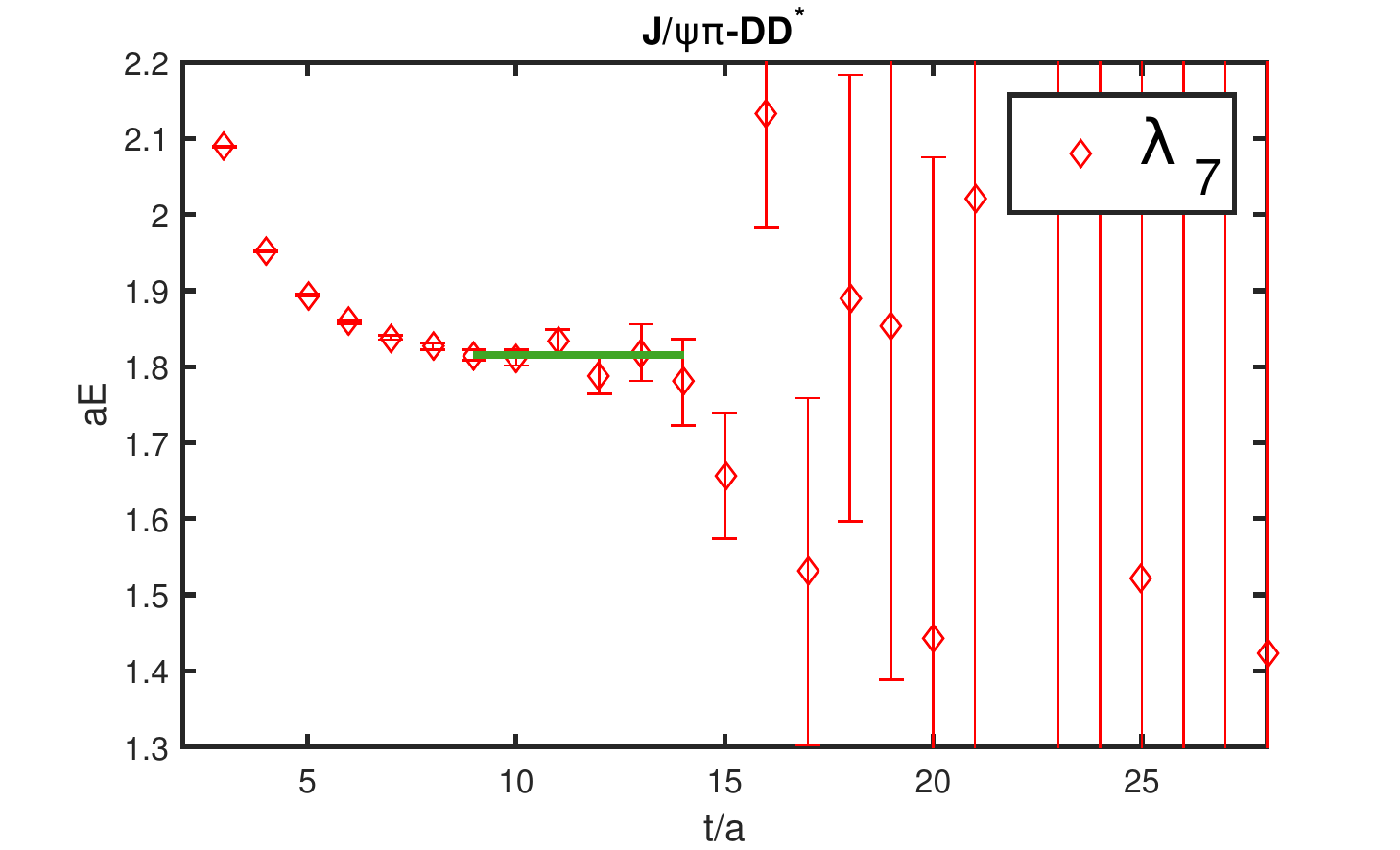}}}
 \caption{Effective mass plateaus for the eight energy levels.
 \label{fig:levels}}%
  \end{figure*}

 All effective mass plateaus are shown in Fig.~\ref{fig:levels} and the
 the final results for the $E_\alpha$'s and the corresponding errors, which are
 analyzed using jackknife method, are
 summarized in Table~\ref{tab:deltaE} together with the fitting ranges and
 $\chi^2$ per degree of freedom.
 It is seen that all of the energy levels develop
 decent plateau behavior.

 \subsection{Extraction of scattering parameters}

 As discussed previously, using Ross-Shaw theory,
 we adopted a four-parameter fit for the data. To be specific,
 we use $M_{11}$, $M_{12}$, $M_{22}$ and $R$ as the parameters to be
 determined, assuming that the effective range are the same in the first
 and second channel. We will collectively denote these four parameters
 as $\{\lambda_i\}$, with $i=1,2,3,4$ corresponding to the above listed
 four parameters.

 Two-channel L\"uscher formula as shown in Eq.~(\ref{eq:luscher_cube_two})
 can be written as,
 \be
 \label{eq:luscher_determine_eq}
 F(\{\lambda_i\};E)=0\;,
 \ee
 for an energy level $E$ in the finite box. The function $F$ involves various
 zeta-functions which can be computed numerically once the parameters are given.
 Thus, for a given set of parameter $\{\lambda_i\}$, the above
 formula can be viewed as an equation for $E$.
 In fact, one can solve for a tower of $E$ which could be compared with what is measured
 from lattice simulations.
 Therefore, following e.g. Ref.~\cite{Dudek:2012gj}, we may construct a $\chi^2$ function as,
 \be
 \label{eq:chi2_def}
 \chi^2(\{\lambda_i\})=\sum_{\alpha\beta}(E^{sol}_\alpha(\{\lambda_i\})-E^{latt}_\alpha)
 \calC^{-1}_{\alpha\beta}(E^{sol}_\beta(\{\lambda_i\})-E^{latt}_\beta)
 \;,
 \ee
 where the summation of $\alpha$ and $\beta$ runs over all
 available energy levels utilized in the fit.
 The energy levels $E^{sol}_\alpha(\{\lambda_i\})$ are obtained by solving
 Eq.~(\ref{eq:luscher_determine_eq}) once the parameters $\{\lambda_i\}$ are given.
 The energy levels $E^{latt}_\alpha$ are the corresponding ones measured from lattice
  data after the GEVP process. The matrix $\calC$ accounts for the correlation
  of these quantities since they are all measured using the same set of ensemble.
 We estimate the covariance matrix $\calC$ from our data using jackknife method.

 \begin{table}
\centering \caption{Parameters $M_{11}$, $M_{22}$, $M_{12}$ and $R$ obtained by
 minimizing the $\chi^2$ function defined in Eq.~(\ref{eq:chi2_def}).
 All quantities are in lattice units.
 The corresponding values of total $\chi^2$ and the number of degrees of freedom are also listed.
 The errors for various parameters are obtained by doing a jackknife analysis.  \label{tab:results_MR}}
\begin{tabular}{c|l|l|l|l|c}
\hline
 No. of levels &$M_{11}$ & $M_{12}$ & $M_{22}$  & $R$ & $\chi^2/N_{d.o.f}$ \\
\hline
 7&  $-7.7(1.8)$ & $3.7(1.1)$ & $-3.3(2.8)$ & $0.12(1.2)$ & $0.57/3$\\
\hline
 7& -7.51(0.45) & 3.7(1.2) & -3.1(1.5) &  &  0.58/4\\
 \hline\hline
\end{tabular}
\end{table}

 After minimizing the function $\chi^2(\{\lambda_i\})$ numerically
 with respect to the four parameters, the optimal values for $\{\lambda_i\}$
 that minimize the $\chi^2$ can be obtained.
 In our simulation, we have obtained 8 energy levels. Since the highest energy
 level is subject to higher states contaminations, we decided to
 fit for the scattering parameters both with and without this level included.
 It turns out that, with the highest energy level included, we keep getting rather
 large $\chi^2$ values while using the lowest $7$ levels yields a tolerable $\chi^2$ value.
 We therefore will only list the best fitted parameter values using only $7$ levels.
 When performing the fit, we could use the the four parameter fit with $R$ included, or
 in the so-called zero range approximation which sets $R=0$.
 It is observed that the four parameter fit yields a value of $R$ that is consistent
 with zero within errors. Therefore, it makes sense to perform the fit within the
 zero range approximation.

 The covariance  matrix of the fitted parameters can be estimated following a standard jackknife procedure.
 For the case of three parameter fit, i.e. in the zero range approximation,
 we obtain the inverse covariance matrix $C^{-1}$ in lattice unit as,
 \be
 \label{eq:Cinv_latt}
 C^{-1}=\left[
 \begin{array}{ccc}
  6.49951962 &  4.3268455 &3.24808892  \\
    4.3268455  &12.51156448 & 9.33781593 \\
  3.24808892 & 9.33781593 & 7.4080769
  \end{array}\right]\;,
 \ee
 where the column and rows of the matrix are labeled according to $(\eta_1,\eta_2,\eta_3)=(M_{11},M_{12},M_{22})$.
 The covariance matrix $C$ itself can be easily obtained as well. If we are to transform these
 matrices to the unit in which $k_{10}=1$, they have to be either multiplied/divided by
 the value of $k^2_{10}$ in lattice units.

 Results of the fitting procedure are tabulated in Table~\ref{tab:results_MR}
 where all dimensional quantities are in lattice units.
 The errors of each individual parameter is the square root of the
 corresponding matrix element listed the covariance matrix $C$.


\subsection{Discussion of the results}
 \label{subsec:confidence_level}

 Since the presence of the effective range parameter $R$ is marginal,
 in the following discussion we will only focus on the case of three parameters.
 For later convenience, we collectively denote these
 parameters as $\eta=(M_{11},M_{12},M_{22})^T\in\mathbb{R}^3$ and the value of $\eta$
 which minimizes the $\chi^2$ function will be denoted as $\eta^*$.
 We will be showing these parameters in units of $k_{10}$.

 In a small neighborhood around the best fitted value for $\eta$ (in some unit),
 the function $\chi^2(\eta)$ can be parameterized as,
 \be
 \label{eq:chi2_confidence_level}
 \chi^2(\eta)=\chi^2(\eta^*)+\frac{1}{2}w^T C^{-1}\cdot w
 \;,
 \ee
 where $w=\eta-\eta^*$ and the matrix $C$ is the covariance matrix for the parameters which also
 yields the errors (and the cross-variance) for the fitted parameters.

 In any case, the matrix $C^{-1}$ is a $3\times 3$ positive-definite symmetric real matrix
 which can be diagonalized via some rotation matrix $R$. In fact, setting
 $x=R\cdot w$ and demanding that the new matrix $R^TC^{-1}R$ being diagonal,
 we have
 \be
 \label{eq:half_majors}
 R^T\cdot C^{-1} R =\mbox{Diag}(1/\sigma^2_1,1/\sigma^2_2,1/\sigma^2_3)
 \;,
 \ee

 Confidence levels can then be set by using the change in the value
 of $\chi^2$ function relative to its minimum value in the parameter space.
 Denoting
 \be
 \Delta\chi^2(\eta)=\chi^2(\eta)-\chi^2(\eta^*)\;,
 \ee
 this quantity in terms of the rotated parameters $x$  is diagonal:
 \be
 \Delta\chi^2(x)=\frac{1}{2}\sum^3_{i=1}\frac{x^2_i}{\sigma^2_i}
 \;.
 \ee
 Therefore, for a given value of $\Delta\chi^2(x)$, the above equation
 becomes a three-dimensional ellipsoid centered at the origin  of $x\in\mathbb{R}^3$ with
 three half-major axis given by: $\sqrt{2\Delta\chi^2}\sigma_1$,
 $\sqrt{2\Delta\chi^2}\sigma_2$ and $\sqrt{2\Delta\chi^2}\sigma_3$, respectively.
 In terms of the original variable $w$, this is a rotated ellipsoid
 with the rotation characterized by the matrix $R$ which diagonalize
 the matrix $C^{-1}$.

\begin{figure*}[htb]
  {\resizebox{0.32\textwidth}{!}{\includegraphics{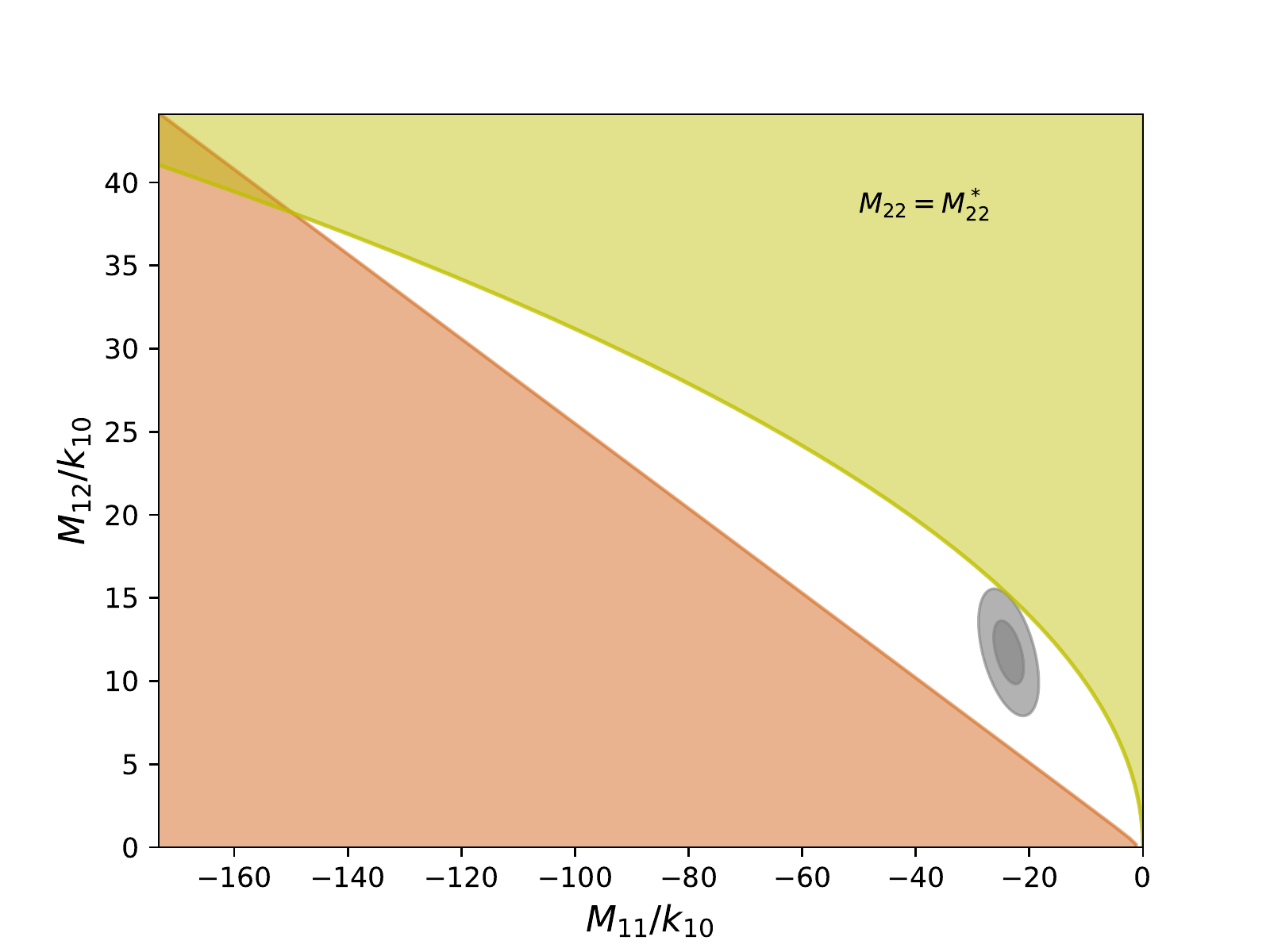}}}
   {\resizebox{0.32\textwidth}{!}{\includegraphics{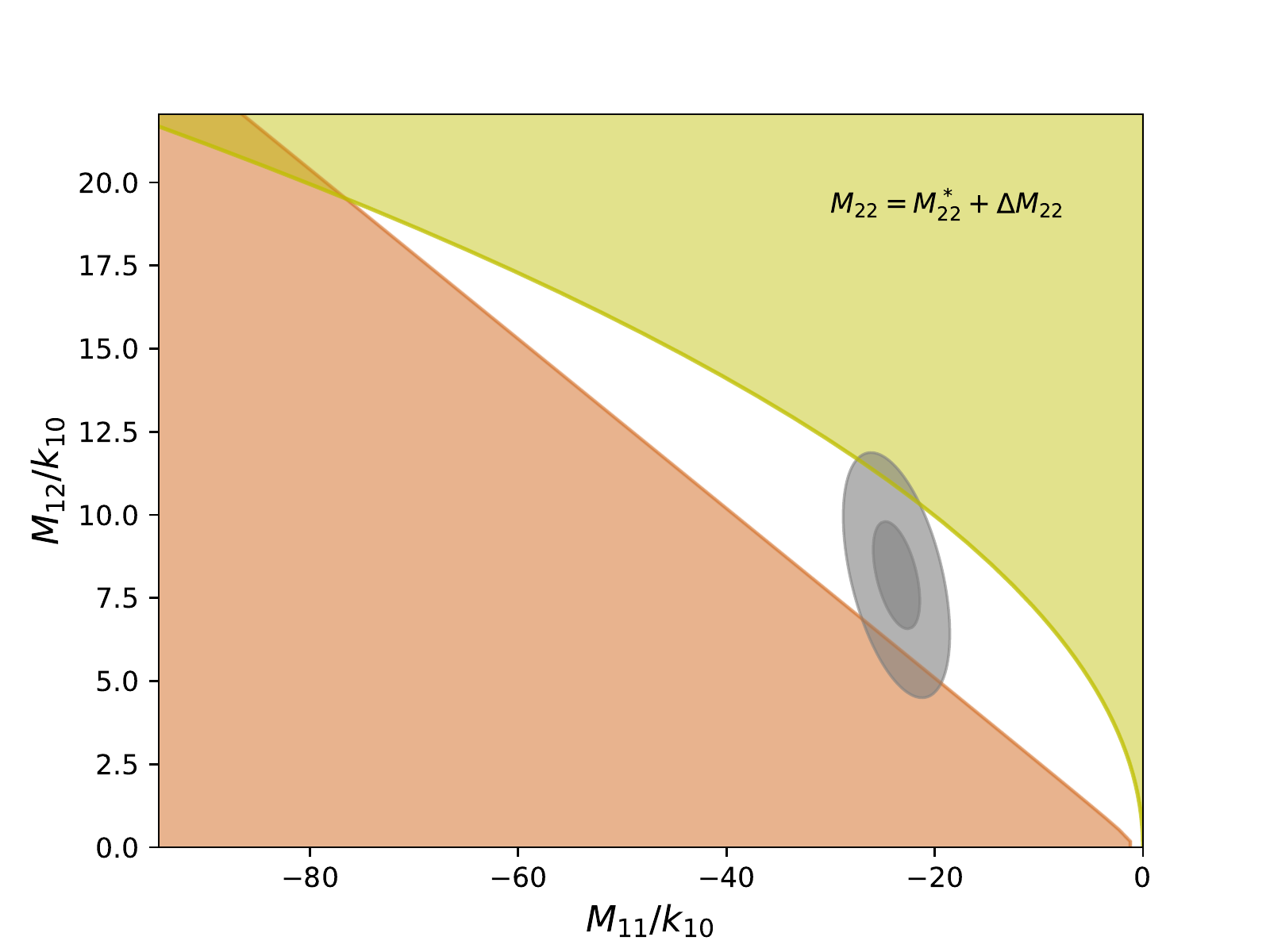}}}
    {\resizebox{0.32\textwidth}{!}{\includegraphics{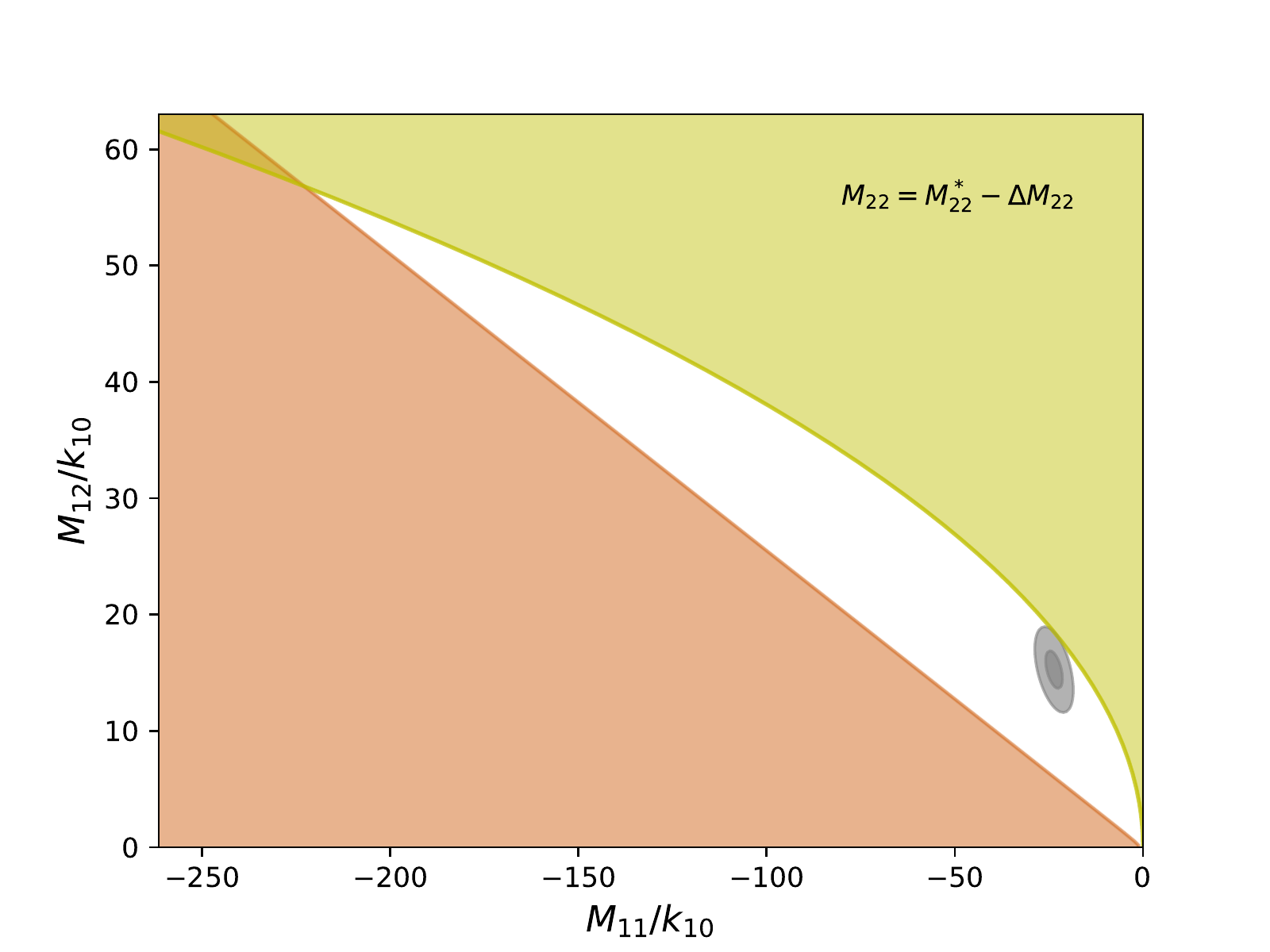}}}
 \caption{Various contours obtained from Eq.~(\ref{eq:ellipse}) together with
 the constraint in Eq.~(\ref{eq:hyperbola})
  and Eq.~(\ref{eq:parabola}). Three different panels, from left to right corresponds
  to different values of $\eta_3\equiv M_{22}$: $M_{22}=M^*_{22}$, $M_{22}=M^*_{22}+\Delta M_{22}$
  and   $M_{22}=M^*_{22}-\Delta M_{22}$, respectively with $\Delta M_{22}$ being the error for $M_{22}$.
  In each panel, the central point of the two ellipses
  corresponds to the best fitted value of $(\eta^*_1,\eta^*_2)$ for that particular value
  of $\eta_3$. The two elliptical shaded region around the central point correspond
  to $1\sigma$ and $3\sigma$ contours
  for the parameters $(\eta_1, \eta_2)=(M_{11},M_{12})$.
  The lower left shaded region corresponds to the narrow resonance condition
  described by inequality~(\ref{eq:hyperbola}) while the upper right shaded
  region corresponds to the fact that the resonance is close to the threshold,
  i.e. inequality~(\ref{eq:parabola}).
  The values for the two ratios have assumed their ``true'' values for $Z_c(3900)$,
  namely:  $R^{\mbox{cut}}_{\mbox{close}}=0.0211$
  and $R^{\mbox{cut}}_{\mbox{narrow}}=0.065$.
 \label{fig:contours1}}%
  \end{figure*}
  It is known that, for three-parameter $\chi^2$ fit, the $1\sigma$, $2\sigma$ and $3\sigma$
 contours
 \footnote{In three dimensions, they are in fact surfaces instead of contours. But since we
 will be showing two-dimensional intersections, we will call them contours instead.}
  can be set by demanding $\Delta\chi^2=3.53, 8.02, 14.2$, respectively.
 Thus, denoting $w=(M_{11}-M^*_{11}, M_{12}-M^*_{12}, M_{22}-M^*_{22})^T$, the ellipsoid,
 \be
 \Delta\chi^2={1\over 2} w^T\cdot C^{-1} \cdot w\;,
 \ee
 centered at $\eta^*=(M^*_{11}, M^*_{12},M^*_{22})$ will enclose the $68$\%, $95.4$\%
 and $99.7$\% probability as far as the parameters are concerned.

\begin{figure*}[htb]
  {\resizebox{0.32\textwidth}{!}{\includegraphics{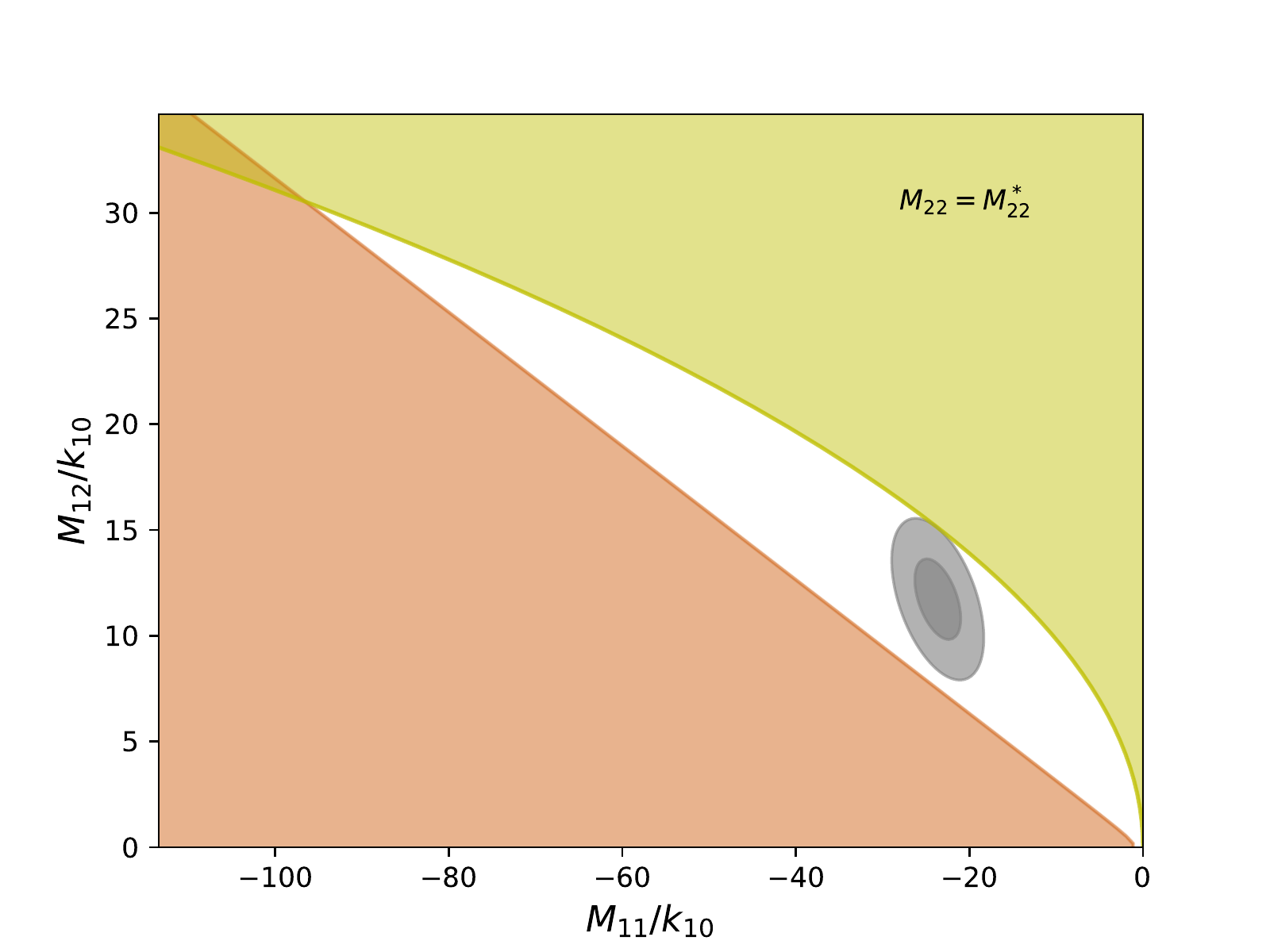}}}
   {\resizebox{0.32\textwidth}{!}{\includegraphics{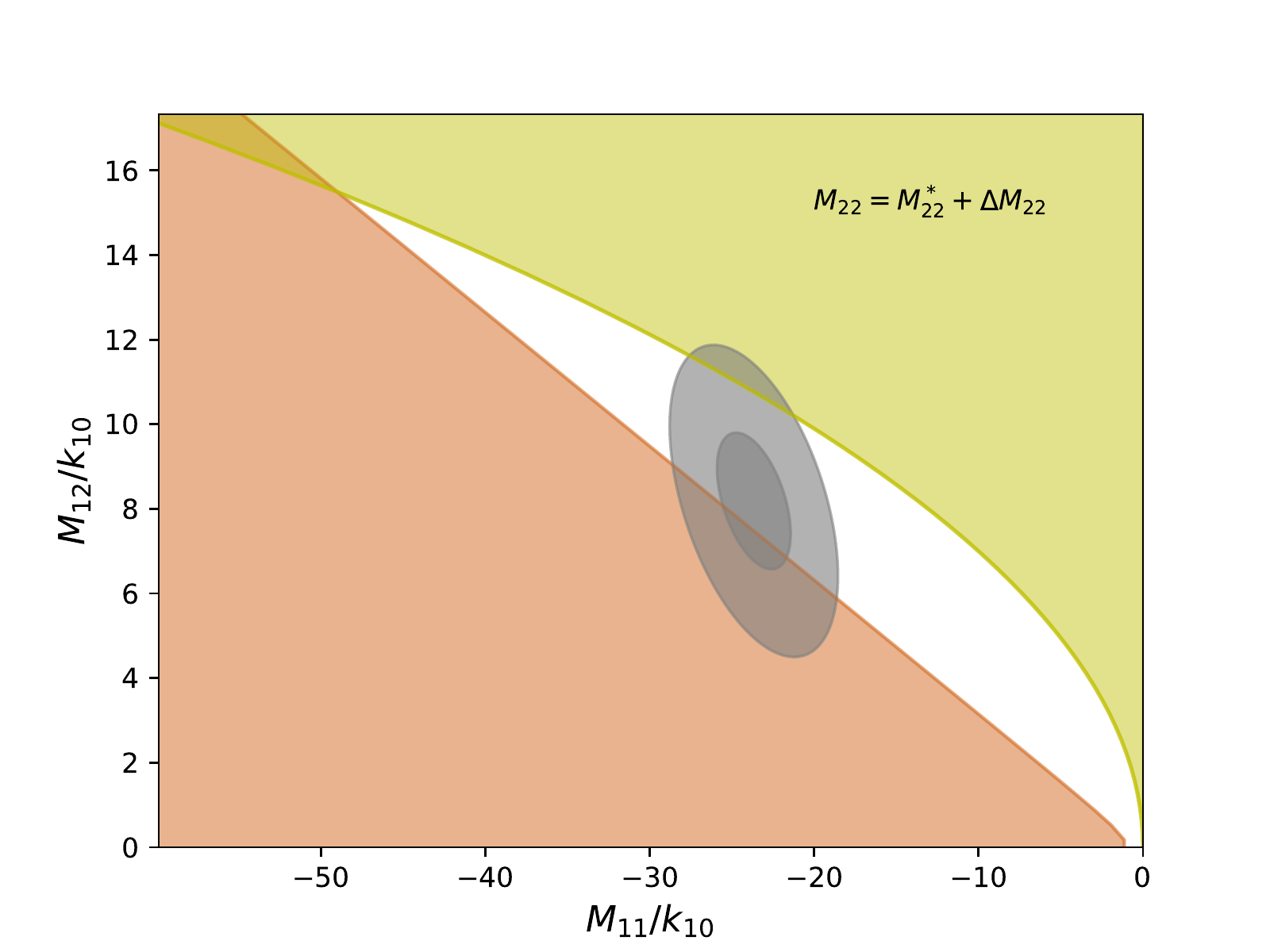}}}
    {\resizebox{0.32\textwidth}{!}{\includegraphics{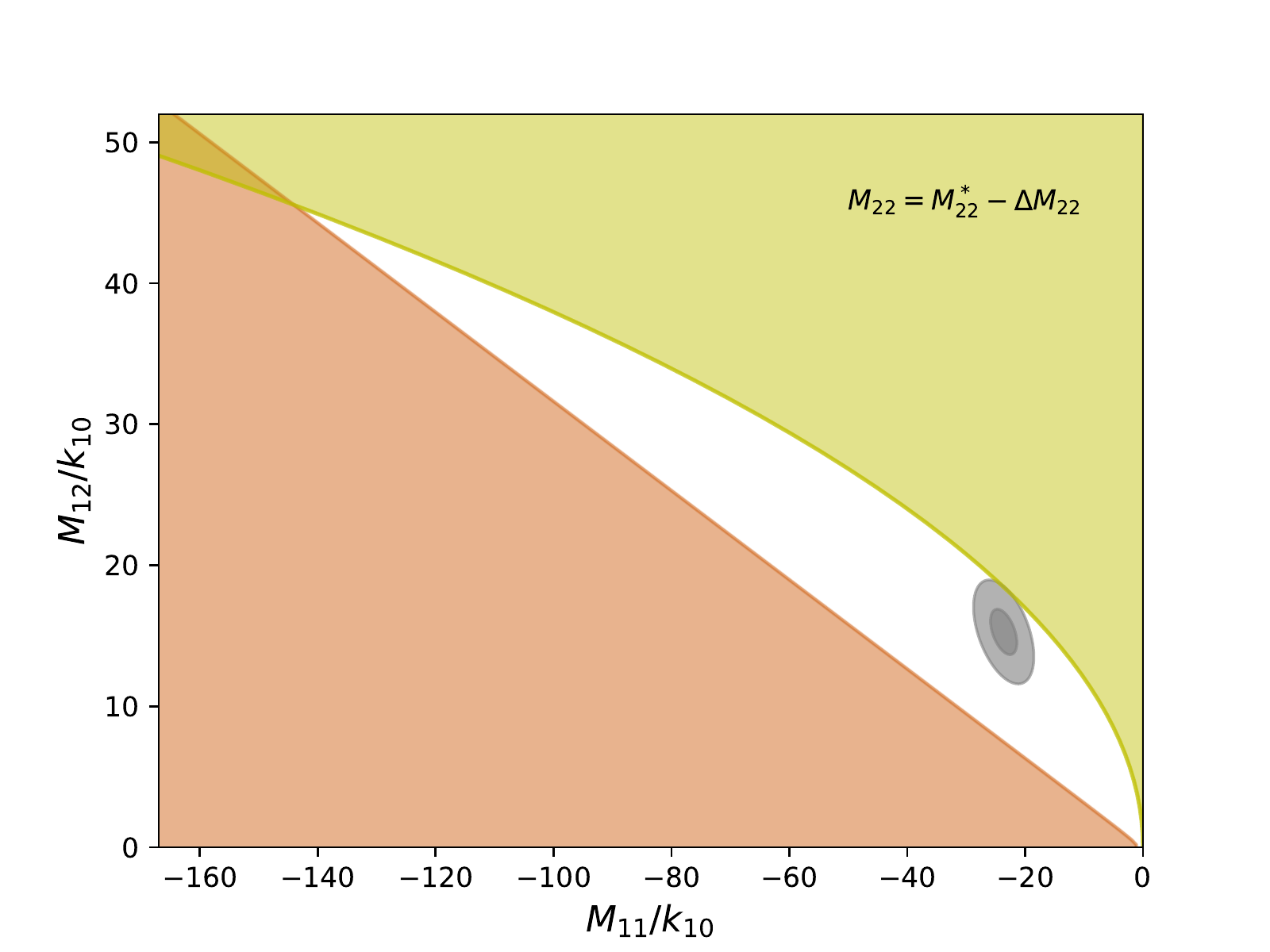}}}
 \caption{Same as the previous figure except that
  the values for the ratios are:  $R^{\mbox{cut}}_{\mbox{close}}=R^{\mbox{cut}}_{\mbox{narrow}}=0.1$.
 \label{fig:contours2}}%
  \end{figure*}
 On the other hand, the closeness and narrowness
 condition given in Eq.~(\ref{eq:close_narrow}) yields
 a parabola and a hyperbola on the $\eta_1-\eta_2$ plane for a given value of $\eta_3=M_{22}$, respectively.
 Therefore, if we demand that $R_{\mbox{close}}$ and $R_{\mbox{narrow}}$ to
 be smaller than some prescribed cut values,
 \be
 R_{\mbox{narrow}} \le R^{\mbox{cut}}_{\mbox{narrow}}\;,
 \;\;
 R_{\mbox{close}} \le R^{\mbox{cut}}_{\mbox{close}}\;.
 \ee
 We can therefore check whether these conditions are supported by our $\chi^2$ fit.

 To summarize, these conditions are, using the unit system in which $k_{10}=1$:
  \ba
 \label{eq:ellipse}
 \frac{1}{2}(\eta-\eta^*)^TC^{-1}\cdot (\eta-\eta^*) &\le & \Delta\chi^2
 \;,
 \\
  \label{eq:hyperbola}
 \eta^2_1-
 \frac{\eta^2_2}{R^{\mbox{cut}}_{\mbox{narrow}} } &\le & 1\;,
 \\
  \label{eq:parabola}
 \eta_1[\eta_3+ R^{\mbox{cut}}_{\mbox{close}} ] &\le &\eta^2_2
 \;.
 \ea
 For a given value of $\Delta\chi^2$, the first inequality~(\ref{eq:ellipse}) implements
 an ellipsoid that encloses a certain probability
 around the best fitted value at $\eta=\eta^*$; the second inequality~(\ref{eq:hyperbola})
 implies a region bounded by a hyperbola in the $\eta_1\eta_2$ plane, independent
 of the value of $\eta_3=M_{22}$; the third inequality, however, requires that
 the point $(\eta_1,\eta_2$ to be in a region that is above a parabola
 for a given value of $\eta_3=M_{22}$.

 Therefore, if we want to have a narrow resonance close enough to
 the threshold, described by two prescribed ratios: $R^{\mbox{cut}}_{\mbox{narrow}}$
 and $R^{\mbox{cut}}_{\mbox{close}}$, these points have to lie within the overlapping regions
 of the above mentioned parabola and hyperbola. By inspecting the location of the overlapping region
 with respect to various constant $\Delta\chi^2$ contours, we are then able to set the
 confidence intervals for the parameters. This could be done when a particular
 value of $\eta_3=M_{22}$ is given.

 In Fig.~\ref{fig:contours1}, the situation is illustrated at three particular
 values (from left to right) of $M_{22}$, namely $M_{22}=M^*_{22}, M^*_{22}+\Delta M_{22}, M^*_{22}-\Delta M_{22}$,
 with $\Delta M_{22}$ being the error for $M_{22}$. Here, all quantities have been converted
 to the unit system in which $k_{10}=1$.
 In each panel of the figure, the common center of the two ellipses
 corresponds to minimum of the $\chi^2$ function, i.e. the best fitted value
 of $(\eta^*_1,\eta^*_2)$ for that particular value of $\eta_3$.
 The two elliptical shaded regions around the common center correspond
 to the $1\sigma$ (68\% probability) and $3\sigma$ (99.7\% probability) contours
 for the parameters $(\eta_1, \eta_2)=(M_{11},M_{12})$.
 The lower left shaded region corresponds to the narrow resonance condition
 described by inequality~(\ref{eq:hyperbola}) while the upper right shaded
 region corresponds to the fact that the resonance is close to the threshold,
 i.e. inequality~(\ref{eq:parabola}).
 In this figure, the values for the two ratios have assumed their ``true''
 physical values for $Z_c(3900)$, i.e. $R^{\mbox{cut}}_{\mbox{close}}=0.0211$
 and $R^{\mbox{cut}}_{\mbox{narrow}}=0.065$.
 It is clearly seen that in this case, the overlapping regions,
 which are located in the top left corner in each panel,
 are very far away from the $3\sigma$ contours. Therefore,
 for this set of parameters, it is highly unlikely that
 the three-parameter Ross-Shaw model can describe a resonance
 that, like the $Z_c(3900)$, is both narrow and close to threshold.

 Since in our simulations we do not have a physical pion mass and
 a physical charm quark mass, the two ratios and also the parameter $k_{10}$
 could change to some values that are not their real world values.
 However, since the ratios are dimensionless, we do not expect them
 to change drastically. Similarly, as we see previously, the value of $k_{10}$
 is also not very different from the physical value. Nevertheless,
 we still take $R^{\mbox{cut}}_{\mbox{narrow}}=R^{\mbox{cut}}_{\mbox{close}}=0.1$
 and inspect the relation of the overlapping regions and the constant
 $\Delta\chi^2$ contours again. These are shown in Fig.~\ref{fig:contours2}.
 It is seen that, even at this set of parameters, the overlapping regions
 are still far from the $3\sigma$ contours, showing that the three-parameter
 Ross-Shaw model can not explain a narrow resonance close to the resonance
 that is described by  $R^{\mbox{cut}}_{\mbox{narrow}}=R^{\mbox{cut}}_{\mbox{close}}=0.1$.

 We have also tried other parameter values and the outcome is quite similar.
 Therefore, as far as the parameters are concerned, it seems that
 the three-parameter Ross-Show model cannot realize a scenario in which
 a narrow resonance appears close enough to the threshold.
 This in fact can be understood from physical arguments as follows:
 The reason is that, the best fitted values for the $M$ matrix elements
 are all very large, either in lattice unit or in $k_{10}$ unit.
 Recall that this matrix is related to $\cot\delta$ matrix, or the
 inverse scattering length matrix of the scattering, c.f. Eq.~(\ref{eq:M_def}).
 Large matrix elements of $M$, if the matrix itself is nonsingular, yields large
 inverse scattering length, meaning a negligible scattering effect.
 This is the reason that the zero-range Ross-Shaw theory had a hard time
 to generate a narrow resonance peak near the threshold.
 Furthermore, if we ask why on earth that the matrix elements of $M$ turn out
 to be large? This is in fact implicitly hidden in our energy levels $E_\alpha$,
 see Table~\ref{tab:deltaE}.
 All energy levels we utilized in the L\"uscher formula are
 rather close to the corresponding free two-particle energy levels.
 This fact in turn generates large values for the $M$ matrix elements.

 However, it is still premature to draw the conclusion
 that the  three-parameter Ross-Shaw theory cannot describe a narrow resonance close to
 the threshold due to the following reasons: In the above arguments we have
 not taken into account the systematic errors. Only statistical errors
 are considered and they are assumed to be normally distributed.
 Although we used dimensionsless quantitites in our study to bypass scale setting problems, there are still several systematic effects. The result was calculated on one nonphysical ensemble. A further study is required to perform an extrapolation to the physical point.
 Another systematic effects is that we have only considered two channels.
 Of course, one could try to add more channels e.g. the $\rho\eta_c$, to the discussion.
 For that purpose, one needs to have much more energy levels since
 a three-channel Ross-Shaw theory, even within the zero range approximation,
 needs $6$ parameters. In order to nail down these, one needs more energy levels.
 This is also something that could be attempted in the future.

 \section{Conclusions}
 \label{sec:conclude}

 Let us now outline the main conclusions from our study:
 \begin{itemize}
 \item In this exploratory lattice study, we utilize the coupled channel
 L\"uscher formula together with the Ross-Shaw theory to
 study the near threshold scattering of $D\bar{D}^*$  which is relevant
 for the exotic state of $Z_c(3900)$.

 \item We single out the most strongly coupled two channels of the problem, namely
 $D\bar{D}^*$ and $J/\psi\pi$. The fact that these two particular channels show strongest coupling
 is supported both by our correlation matrix estimation and by the experimental facts.

 \item Our results show that the inverse scattering length parameters $M_{11}$,
 $M_{12}$ and $M_{22}$ are huge in magnitude,
 indicating that it is unlikely to generate any resonance behavior
 that is both narrow and close enough to the threshold.

 \item Unlike what the HALQCD collaboration finds out, our results do not support
 a narrow resonance-like peak close to the threshold by taking into account
 the most relevant two coupled channels in the problem.

 \item However, one has to keep in mind that we have not
 estimated  the systematic uncertainties. All error estimates are purely
 statistical and the systematics could be due to
 finite lattice spacing, non-physical pion and charm quark masses,
 finite volume effects, etc. which needs to be clarified in future studies.

 \end{itemize}

 To summarize, in this paper we present an exploratory lattice study
 for the coupled channel scattering near the $D\bar{D}^*$ threshold using coupled-channel
 L\"uscher formalism. We utilize $2+1+1$ twisted mass fermion configurations at
 a lattice spacing of $0.0863$fm with the pion mass value of about $320$MeV.
 The most two relevant channels, namely $J\psi\pi$ and $D\bar{D}^*$ are studied,
 which are singled out from four channels by a correlation matrix analysis.
 To extract the scattering information, we fit our lattice results using
 the Ross-Shaw theory, a multi-channel generalization of the conventional
 effective range theory.
 Using our lattice data, the matrix elements of $M$ matrix are obtained together
 with the effective range parameter, although the latter turns out to be marginal.

 Our results indicate that, it is unlikely to satisfy both narrow resonance  condition
 and the condition that is close enough to the threshold, unless the parameters
 happen to reside in a small corner  far away from the best fitted values
 in our  parameter space.
 Keep in mind that we have only considered the statistical errors,
 further studies with more lattice spacings and volumes, pion masses and charm quark masses,
 or even more channels are needed to quantify these systematic effects.
 We hope that this exploratory study will shed some light on the
 multi-channel study of charmed meson scattering which is
 intimately related to $Z_c(3900)$.

 \section*{Acknowledgments}

 The authors would like to thank F.~K. Guo, U.~Meissner, A.~Rusetsky, C.~Urbach for helpful discussions.
 This work is supported in part by the Ministry of Science and Technology of
 China (MSTC) under 973 project "Systematic studies on light hadron spectroscopy", No. 2015CB856702.
 It is also supported in part by the DFG and the NSFC through funds
 provided to the Sino-Germen CRC 110 ``Symmetries and the Emergence
 of Structure in QCD'', DFG grant no. TRR~110 and NSFC grant No. 11621131001.
 This work is supported in part by the National Science Foundation of China (NSFC) under the project No. 11775229, 11875169 and by the Youth Innovation Promotion Association of CAS (2015013).
 LL acknowledges the support from the Key Research Program of the Chinese Academy of Sciences, Grant NO. XDPB09.




%

 \end{document}